\documentclass[12pt,a4paper]{article}
\pdfoutput=1
\usepackage[a4paper]{geometry}

\usepackage{amsmath}
\usepackage{amsfonts}
\usepackage{amssymb}
\usepackage{dsfont}
\usepackage[nosort]{cite}
\usepackage{hyperref}
\usepackage[margin=20pt,font=small,labelfont=bf]{caption}
\usepackage{multirow}

\newcommand{\eq}[1]{\begin{equation}
                     \begin{split} #1 \end{split}
                     \end{equation}}
\newcommand{\ov}{\overline}
\newcommand{\op}{\hspace{1pt}}

\newcommand{\bom}[1]{\fboxsep2mm\fbox{$\displaystyle{ #1}$}}

\allowdisplaybreaks[2]
\numberwithin{equation}{section}

\begin{document}

\vspace*{1cm}

\begin{center}
{\LARGE
Moduli stabilization with non-geometric fluxes \\
--- \\
comments on tadpole contributions \\ and 
de-Sitter vacua \\}
\end{center}

\vspace{0.5cm}

\begin{center}
  Erik Plauschinn
\end{center}

\vspace{0.4cm}

\begin{center} 
\textit{
Institute for Theoretical Physics, Utrecht University \\
Princetonplein 5, 3584CE Utrecht \\
The Netherlands \\
}
\end{center} 

\vspace{2cm}

%%%%%%%%%%%%%%%%%%%%%%%%%%%%%%%%%%%%%%%%%%%%%%%
%%%%%%%%%%%%%%%%%%%%%%%%%%%%%%%%%%%%%%%%%%%%%%%
%%%%%%%%%%%%%%%%%%%%%%%%%%%%%%%%%%%%%%%%%%%%%%%
%%%%%%%%%%%%%%%%%%%%%%%%%%%%%%%%%%%%%%%%%%%%%%%

\begin{abstract}
\noindent
We study moduli stabilization for type IIB orientifold compactifications 
on  Calabi-Yau three-folds with \mbox{(non-)}geo\-metric fluxes.
For this setting it is possible to stabilize all closed-string moduli classically without
the need for non-perturbative contributions, and  examples 
of stable de-Sitter constructions can be found in the literature which violate a prominent swampland conjecture. 

\noindent
In this paper we derive general properties of non-geometric flux-com\-pac\-ti\-fi\-cations, we argue that 
the contribution of fluxes to the tadpole cancellation conditions should be 
similar to D-branes (and not anti-D-branes), and we exclude supersymmetric 
Minkowski vacua for certain cases. We also reassess known 
stable de-Sitter constructions 
with \mbox{(non-)}geometric $H$-, $F$-, $Q$- and $R$-fluxes 
and argue that these are not consistent in string theory.

\end{abstract}

%%%%%%%%%%%%%%%%%%%%%%%%%%%%%%%%%%%%%%%%%%%%%%%
%%%%%%%%%%%%%%%%%%%%%%%%%%%%%%%%%%%%%%%%%%%%%%%
%%%%%%%%%%%%%%%%%%%%%%%%%%%%%%%%%%%%%%%%%%%%%%%
%%%%%%%%%%%%%%%%%%%%%%%%%%%%%%%%%%%%%%%%%%%%%%%

\clearpage

\tableofcontents

%%%%%%%%%%%%%%%%%%%%%%%%%%%%%%%%%%%%%%%%%%%%%%%
%%%%%%%%%%%%%%%%%%%%%%%%%%%%%%%%%%%%%%%%%%%%%%%
%%%%%%%%%%%%%%%%%%%%%%%%%%%%%%%%%%%%%%%%%%%%%%%
%%%%%%%%%%%%%%%%%%%%%%%%%%%%%%%%%%%%%%%%%%%%%%%
%%%%%%%%%%%%%%%%%%%%%%%%%%%%%%%%%%%%%%%%%%%%%%%
%%%%%%%%%%%%%%%%%%%%%%%%%%%%%%%%%%%%%%%%%%%%%%%
%%%%%%%%%%%%%%%%%%%%%%%%%%%%%%%%%%%%%%%%%%%%%%%
%%%%%%%%%%%%%%%%%%%%%%%%%%%%%%%%%%%%%%%%%%%%%%%
%%%%%%%%%%%%%%%%%%%%%%%%%%%%%%%%%%%%%%%%%%%%%%%
%%%%%%%%%%%%%%%%%%%%%%%%%%%%%%%%%%%%%%%%%%%%%%%
%%%%%%%%%%%%%%%%%%%%%%%%%%%%%%%%%%%%%%%%%%%%%%%
%%%%%%%%%%%%%%%%%%%%%%%%%%%%%%%%%%%%%%%%%%%%%%%

\section{Introduction}

Constructing  stable de-Sitter vacua in string theory is  notoriously difficult. 
An overview of existing de-Sitter constructions 
and discussions of their validity can be found in the reviews \cite{Danielsson:2018ztv} 
and \cite{Palti:2019pca}, but to the best of our knowledge no 
generally-accepted, fully-consistent 
and explicitly worked-out example of a stable de-Sitter vacuum in string theory is known.
This observation may have led the authors of \cite{Obied:2018sgi,Garg:2018reu,Ooguri:2018wrx}
to conjecture that de-Sitter vacua cannot be realized in string theory, or any consistent theory of 
quantum gravity.
As a potential counter-example to this conjecture non-geometric flux-compactifications
are often mentioned, which are the topic of this work.

Non-geometric backgrounds are configurations which cannot be 
described in terms of  Riemannian geometry. Such spaces are inconsistent for point particles
but are well-defined for strings, and for these backgrounds 
the transition functions between local charts are required to include
T-duality transformations  \cite{Hull:2004in}.
For a review of this topic we refer to \cite{Plauschinn:2018wbo}. 
The standard example 
for a non-geometric background is obtained by applying successive T-duality transformations to a
three-torus with $H$-flux, leading to a twisted torus with geometric $F$-flux 
\cite{Dasgupta:1999ss,Kachru:2002sk}, a T-fold with non-geometric $Q$-flux \cite{Hull:2004in}
and a non-associative space with a non-geometric $R$-flux \cite{Shelton:2005cf}.
When compactifying string theory in the presence of fluxes, the $H$-flux 
generates the familiar Gukov-Vafa-Witten superpotential \cite{Gukov:1999ya}
in the lower-dimensional theory. Applying then T-duality transformations 
leads to contributions of the geometric $F$- and of the non-geometric $Q$-
and $R$-fluxes \cite{Shelton:2005cf,Villadoro:2006ia,Shelton:2006fd},
but an uplift of these four-dimensional theories to ten-dimension is often not known. 
However, we  note that certain compactifications with geometric and non-geometric fluxes fit into 
the framework of $SU(3)\times SU(3)$-structure compactifications which have been studied in
\cite{Grana:2005ny,Grana:2006hr,Micu:2007rd,Cassani:2007pq}.

Coming back to de-Sitter vacua, for compactifications including positive-ten\-sion objects such as D-branes 
the Maldacena-Nu\~nez no-go theorem excludes the existence of de-Sitter 
vacua \cite{Maldacena:2000mw}.
However, this no-go theorem does  not hold in the presence of orientifold planes or 
for non-geometric backgrounds, 
and the purpose of the present work is to investigate the latter case.
We focus on type IIB  string-theory compactifications on Calabi-Yau orientifolds
with O3-/O7-planes and \mbox{(non-)}geometric $H$-, $F$-, $Q$- and $R$-fluxes
which, in the context of de-Sitter vacua, have been considered before for instance in 
\cite{
deCarlos:2009fq, 
deCarlos:2009qm, 
Dibitetto:2010rg, 
Dibitetto:2011gm,
Danielsson:2012by,
Blaback:2013ht, 
Damian:2013dq,
Damian:2013dwa,  
Blaback:2015zra,
Shukla:2016xdy,
Cribiori:2019hrb,
CaboBizet:2020cse
}.
Unfortunately, we were neither able to show that 
stable de-Sitter vacua from non-geometric fluxes cannot exists --- nor able 
to construct a consistent de-Sitter model for this setting. Nevertheless, we 
made progress in understanding non-geometric flux vacua and, in particular,
we argue that all stable de-Sitter constructions
with  $H$-, $F$-, $Q$- and $R$-fluxes known to us are inconsistent 
in string theory.
Our findings are reported in the present paper, which 
is organized as follows:
\begin{itemize}

\item In section~\ref{sec_prelim} we review type IIB Calabi-Yau orientifold 
compactifications with O3-/O7-planes and non-geo\-me\-tric 
$H$-, $F$-, $Q$- and $R$-fluxes.
This section contains no new results, but
it clarifies for instance how non-geometric fluxes should be quantized
and it may serve as a modern 
introduction to the topic.

\item In section~\ref{sec_tadpole_charges} we discuss how (non-)geometric 
fluxes contribute to the tadpole cancellation conditions. 
By requiring the rank of the four-dimensional gauge group 
to be bounded (in agreement with \cite{Vafa:2005ui}), we present arguments 
that fluxes should contribute in
the same way as supersymmetric D-branes and not as anti-D-branes. 
This condition
excludes a known de-Sitter construction with 
non-geometric fluxes. 

\item In section~\ref{sec_mod_stab} we study the scalar potential 
induced by non-geometric fluxes. We derive a simple 
expression for the scalar potential at the minimum, and 
we derive the necessary conditions $h^{1,1}_{\vphantom{-}}\leq h^{2,1}_-$ and 
$h^{2,1}_+=0$ on the Hodge numbers for 
stabilization of all closed-string moduli.

\item In section~\ref{sec_minimal_case} we specialize to 
the situation $h^{1,1}_{\vphantom{-}}=h^{2,1}_-$.
We rewrite the scalar potential and eliminate the 
contribution of the Ramond-Ramond three-form flux in favor of 
tadpole charges, 
which is a form suitable for computer-aided scans for vacua. 
We furthermore show that supersymmetric Minkowski vacua 
do not exists and we discuss necessary conditions 
for stable de-Sitter minima.

\item In section~\ref{sec_desitter} we review 
known de-Sitter constructions with non-geometric fluxes.
For the stable models which in addition to the Ramond-Ramond three-form flux only involve
$H$-, $F$-, $Q$- and $R$-fluxes
we conclude  -- using the constraint from 
section~\ref{sec_tadpole_charges} -- that none of them 
are consistent in string theory.

\item In section~\ref{sec_sum} we summarize our findings.

\end{itemize}

%%%%%%%%%%%%%%%%%%%%%%%%%%%%%%%%%%%%%%%%%%%%%%%
%%%%%%%%%%%%%%%%%%%%%%%%%%%%%%%%%%%%%%%%%%%%%%%
%%%%%%%%%%%%%%%%%%%%%%%%%%%%%%%%%%%%%%%%%%%%%%%
%%%%%%%%%%%%%%%%%%%%%%%%%%%%%%%%%%%%%%%%%%%%%%%
%%%%%%%%%%%%%%%%%%%%%%%%%%%%%%%%%%%%%%%%%%%%%%%
%%%%%%%%%%%%%%%%%%%%%%%%%%%%%%%%%%%%%%%%%%%%%%%

\section{Type IIB compactifications with fluxes}
\label{sec_prelim}

We start by introducing our notation and conventions for compactifications of type IIB 
string theory on Calabi-Yau orientifolds with fluxes.
This section contains no new results, and for the reader familiar with the topic it is safe to 
only skim over the formulas and move to the next section.

%%%%%%%%%%%%%%%%%%%%%%%%%%%%%%%%%%%%%%%%%%%%%%%
%%%%%%%%%%%%%%%%%%%%%%%%%%%%%%%%%%%%%%%%%%%%%%%
%%%%%%%%%%%%%%%%%%%%%%%%%%%%%%%%%%%%%%%%%%%%%%%
%%%%%%%%%%%%%%%%%%%%%%%%%%%%%%%%%%%%%%%%%%%%%%%

\subsection{Type IIB orientifolds with O3-/O7-planes}
\label{sec_prelim_set}

In this subsection we summarize some properties of type IIB orientifold compactifications
with O3- and O7-planes which are relevant for our subsequent discussion. 

%%%%%%%%%%%%%%%%%%%%%%%%%%%%%%%%%%%%%%%%%%%%%%%
%%%%%%%%%%%%%%%%%%%%%%%%%%%%%%%%%%%%%%%%%%%%%%%

\subsubsection*{Cohomology}

We consider type IIB string theory on $\mathbb R^{3,1}\times \mathcal X$
subject to an orientifold projection of the form $(-1)^{F_{\rm L}} \Omega_{\rm P}\op\sigma$.
$F_{\rm L}$ denotes the left-moving fermion number and $\Omega_{\rm P}$ is the world-sheet 
parity operator, and their action on the type IIB field content can be found for instance in 
\cite{Plauschinn:2009izh}. 
The compact space $\mathcal X$ is chosen to be a Calabi-Yau three-fold, 
and we impose a holomorphic involution $\sigma$ 
on $\mathcal X$ such that 
its K\"ahler  and holomorphic three-form   satisfy
$\sigma^*J=+J$ and $\sigma^*\op\Omega=-\Omega$.
The fixed loci of $\sigma$ in $\mathbb R^{3,1}\times \mathcal X$ 
correspond to O3- and O7-planes.

Since $\sigma$ is an involution, the cohomology groups of $\mathcal X$ split into 
even and odd eigenspaces as
$H^{p,q}(\mathcal X) = H^{p,q}_+(\mathcal X) \oplus H^{p,q}_-(\mathcal X)$
with dimensions
$h^{p,q} = h^{p,q}_+ + h^{p,q}_-$.
For the even cohomology groups  of $\mathcal X$ we
introduce bases in the following way
\eq{
  \label{basis_002}
  \arraycolsep2pt
  \begin{array}{lcl@{\hspace{30pt}}lcl@{\hspace{60pt}}lcl}
  \omega_0& \in & H^{3,3}_+(\mathcal X) \,,
  \\[2pt]
  \omega_{a} &\in& H^{1,1}_+ (\mathcal X) \,,
  & \omega_{\alpha} &\in&H^{1,1}_- (\mathcal X) \,,
  & a &=&1,\ldots, h^{1,1}_+\,,
  \\[2pt]
  \sigma^{a} &\in& H^{2,2}_+ (\mathcal X) \,,
  & \sigma^{\alpha} &\in&H^{2,2}_- (\mathcal X) \,,
  & \alpha &=&1,\ldots, h^{1,1}_-\,,
  \\[2pt]
  \sigma^0 &\in & H^{0,0}_+ (\mathcal X) \,,
  \end{array}
}
which can be chosen such that  the only non-vanishing pairings satisfy
the relations
$\int_{\mathcal X} \omega_0\wedge \sigma^0 = 1$,
$\int_{\mathcal X} \omega_a\wedge \sigma^b = 1$ and
$\int_{\mathcal X} \omega_{\alpha}\wedge \sigma^{\beta} = \delta_{\alpha}{}^{\beta}$.
For the third cohomology group
we introduce symplectic bases as
\eq{
  \label{basis_001}
  \arraycolsep2pt
  \begin{array}{lcl@{\hspace{60pt}}lcl}
  \{\alpha_I,\beta^I\}&\in& H^3_-(\mathcal X)\,,
  & I &=& 0,\ldots, h^{2,1}_-\,,
  \\[2pt]
 \{\alpha_{\lambda},\beta^{\lambda}\} &\in& H^3_+(\mathcal X)\,,
  & \lambda &=& 1,\ldots, h^{2,1}_+\,,
 \end{array}
 }
with the only non-vanishing pairings 
$\int_{\mathcal X} \alpha_I \wedge \beta^J =\delta_I{}^J$ and 
$\int_{\mathcal X} \alpha_{\kappa} \wedge \beta^{\lambda} =\delta_{\kappa}{}^{\lambda}$.

%%%%%%%%%%%%%%%%%%%%%%%%%%%%%%%%%%%%%%%%%%%%%%%
%%%%%%%%%%%%%%%%%%%%%%%%%%%%%%%%%%%%%%%%%%%%%%%

\subsubsection*{Moduli}

The effective  theory obtained after compactification
can be described in terms of four-dimensional $\mathcal N=1$ supergravity.
The bosonic field content originating from the closed-string sector
is summarized in table~\ref{lu_007}, and for details we refer to \cite{Grimm:2004uq}.
%%%%%%%%%%%%%%%
%%%%%%%%%%%%%%%
\begin{table}[t]
\centering
\renewcommand{\arraystretch}{1.1}
\tabcolsep10pt
\begin{tabular}{lcc}
  \mbox{fields} & \mbox{multiplicity} & \mbox{symbol} \\
  \hline\hline
  \mbox{four-dimensional metric} & $1$ & $g$  \\
  \mbox{$U(1)$ vector fields} & $h^{2,1}_+$ & $U^{\lambda}$  \\
  \mbox{complex-structure moduli} & $h^{2,1}_-$ & $z^i$ \\
  \mbox{axio-dilaton} & $1$ & $\tau$\\
  \mbox{K\"ahler moduli} & $h^{1,1}_+$ & $T_a$  \\
  \mbox{axionic moduli} & $h^{1,1}_-$& $G^{\alpha}$ 
\end{tabular}
\caption{Bosonic massless field content 
corresponding to the closed-string sector
after compactifying
type IIB string theory on Calabi-Yau orientifolds
with O3-/O7-planes.\label{lu_007}}
\end{table}
%%%%%%%%%%%%%%%
%%%%%%%%%%%%%%%
The complex-structure moduli $z^i$ with $i=1,\ldots, h^{2,1}_-$ are encoded in the 
holomorphic three-form $\Omega$, 
while the axio-dilaton, K\"ahler moduli and axionic moduli are contained in 
the multiform \cite{Benmachiche:2006df}
\eq{
  \label{hi_071}
  \Phi = e^{B} \mathcal C 
   + i \,\mbox{Re} \left( e^{-\phi}\op \lambda \left[ e^{B- i\op J} \right] \right) .
}    
Here, $\mathcal C$ denotes the sum over all  Ramond-Ramond potentials $\mathcal C = \sum_p C_p$, 
$B$ denotes the Kalb-Ramond field and  we introduced
an operator $\lambda$ acting on a $p$-form $A_{(p)}$ as $\lambda[A_{(p)}] = (-1)^{[\frac{p}{4}]} A_{(p)}$
with $[\ldots]$ denoting the integral part. 
Expanding $\Phi$ in the bases \eqref{basis_002} defines the moduli in the following way
\eq{
\label{hi_007}
  \Phi = \tau  + G^{\alpha} \omega_{\alpha} +  T_{a} \op \sigma^{a} \,,
}
and the precise form of $\tau$, $G^{\alpha}$ and $T_a$ 
can be found for instance in \cite{Grimm:2004uq} or by working out 
the definition of $\Phi$. For later convenience we group the K\"ahler-sector moduli together as
\eq{
  \label{lu_001}
  \mathsf T^A = (\op \tau, G^{\alpha}, T_a) \,, \hspace{50pt} \mathsf T^A = c^A + i \op \tau^A \,,
}
where $c^A$ and $\tau^A$ denote the real and imaginary parts of $\mathsf T^A$
and where we introduced a collective index $A=0,\ldots, h^{1,1}$.

%%%%%%%%%%%%%%%%%%%%%%%%%%%%%%%%%%%%%%%%%%%%%%%
%%%%%%%%%%%%%%%%%%%%%%%%%%%%%%%%%%%%%%%%%%%%%%%

\subsubsection*{K\"ahler potential}

The moduli-space geometry is captured by a K\"ahler potential $K$, which
splits into the K\"ahler sector containing the axio-dilaton, K\"ahler moduli and 
axionic moduli and the complex-structure sector  as follows
\eq{
  \label{lu_005}
  K  =  K_{\rm K} + K_{\rm cs} \,,
  \hspace{50pt}
  \arraycolsep2pt
  \begin{array}{lcl}
  K_{\rm K} &=& \displaystyle  -\log\left[\vphantom{\bigl(} - i\op (\tau-\ov \tau)\right]   -2 \log\left[ \mathcal{V} + \frac{\xi}{2} \right],  \\[10pt]
  K_{\rm cs} &=& \displaystyle  -\log\left[ +\op i \int_{\mathcal X}\Omega\wedge \ov \Omega \op\right].
  \end{array}
}

For the K\"ahler sector we note that 
the Einstein-frame volume of the Calabi-Yau three-fold is denoted by $\mathcal V$
which depends implicitly on  
$\tau$, $G^{\alpha}$ and $T_a$, and we included 
$\alpha'$-corrections  encoded in 
\raisebox{0pt}[0pt][0pt]{$\xi = -\frac{\zeta(3)\op\chi(\mathcal X)\op(\tau-\ov\tau)^{3/2}}{2\op(2\pi)^3\op(2i)^{3/2}}$} \cite{Becker:2002nn}. 
In the following we will employ the notation 
\eq{
  \arraycolsep2pt
  \begin{array}{lcl@{\hspace{50pt}}lcl}
  K_A &=& \partial_{\mathsf T^A} K\,, &
  G_{A\ov B} &=& \partial_{\mathsf T^A\vphantom{\ov{\mathsf T}^B}}  \partial_{\ov{\mathsf T}^B}K\,,
  \\[6pt]
  K^A &=& G^{A\ov B} K_{\ov B} \,,
  &
  G^{A\ov B} &\equiv& (G^{-1})^{A\ov B}\,, 
  \end{array}
}
and from \eqref{lu_005} we can determine the relations 
\eq{
  \label{lu_006}
 K^A = - (\mathsf T- \ov{\mathsf{T}})^A\,, 
 \hspace{50pt} K_{A\vphantom{\ov B}}\op G^{A\ov B} K_{\ov B} = 4\,.
}
We emphasize that our  analysis below depends on \eqref{lu_006}, which we have checked explicitly 
for the $\alpha'$-corrected K\"ahler potential $K_{\rm K}$ shown in \eqref{lu_005}. 
However, we do not know wether \eqref{lu_006} also holds for the full quantum expression.

In the complex-structure sector $\Omega$ denotes again the holomorphic three-form of 
$\mathcal X$, and the K\"ahler covariant derivatives $D_i = \partial_{z^i} + 
\partial_{z^i} K$ acting on $\Omega$  provide a basis $\chi_i = D_i \op\Omega$
for $H^{2,1}_-(\mathcal X)$.  
In addition to \eqref{basis_001}, an alternative basis for $H^3(\mathcal X)$ is therefore given 
by $\{\Omega, \chi_{i\vphantom{\ov i}}, \ov \chi_{\ov i}, \ov \Omega\}$
where the bar denotes complex conjugation. A general three-form can then be expanded as
\eq{
  \label{lu_010}
  A = a^0\op \Omega + a^i \chi_{i\vphantom{\ov i}} +  a^{\ov i}\op \ov \chi_{\ov i} + a^{\ov 0}\op \ov\Omega \,,
}
and, with the Hodge-star operator acting as $\star \Omega = - i\op \Omega$ and $\star\chi_i = + i\op \chi_i$,
for two three-forms $A$ and $B$ we obtain the relation
\eq{
\label{lu_009}
a^{\ov 0} b^{0}+ a^{i\vphantom{\ov j}} G_{i \ov j} \op b^{\ov j} 
&= \frac{e^{K_{\rm cs}}}{2}
 \int_{\mathcal X} A\wedge \bigl( \star B - i  B\bigr) \,.
}

%%%%%%%%%%%%%%%%%%%%%%%%%%%%%%%%%%%%%%%%%%%%%%%
%%%%%%%%%%%%%%%%%%%%%%%%%%%%%%%%%%%%%%%%%%%%%%%
%%%%%%%%%%%%%%%%%%%%%%%%%%%%%%%%%%%%%%%%%%%%%%%
%%%%%%%%%%%%%%%%%%%%%%%%%%%%%%%%%%%%%%%%%%%%%%%

\subsection{Fluxes}
\label{sec_flux_intro}

To generate a potential for the moduli in the effective theory
we introduce fluxes, and to preserve four-dimensional Poincar\'e invariance
they are chosen to extend only along the compact space.
The type of fluxes we consider in the
Ramond-Ramond (R-R) and Neveu-Schwarz-Neveu-Schwarz (NS-NS) 
sectors are
\eq{
  \label{fluxes_072}
  \begin{array}{l@{\hspace{4pt}}l@{\hspace{50pt}}l}
  \mbox{R-R} &\mbox{sector:} & F_3 \,, \\[4pt]
  \mbox{NS-NS} &\mbox{sector:} &   H\,, \hspace{4pt} F\,, \hspace{4pt} Q\,,\hspace{4pt} R\,.
  \end{array}
}

%%%%%%%%%%%%%%%%%%%%%%%%%%%%%%%%%%%%%%%%%%%%%%%
%%%%%%%%%%%%%%%%%%%%%%%%%%%%%%%%%%%%%%%%%%%%%%%

\subsubsection*{(Non-)geometric fluxes}

In the NS-NS sector $H$ denotes the familiar $H$-flux, and its T-dual 
completions are the geometric $F$-flux and the non-geometric 
$Q$- and $R$-fluxes \cite{Dasgupta:1999ss,Kachru:2002sk,Hull:2004in,Shelton:2005cf}. 
A detailed discussion of non-geometric fluxes can be found 
for instance in \cite{Plauschinn:2018wbo}, but here we only note
that they can be interpreted as operators acting 
on the cohomology  as
\eq{
  \renewcommand{\arraystretch}{1.2}
  \arraycolsep3pt
  \begin{array}{l@{\hspace{6pt}}c@{\hspace{12pt}}lcl}
  H\,\wedge & :& \mbox{$p$-form} &\to& \mbox{$(p+3)$-form} \,, \\
  F\,\circ & :& \mbox{$p$-form} &\to& \mbox{$(p+1)$-form} \,, \\
  Q\,\bullet & :& \mbox{$p$-form} &\to& \mbox{$(p-1)$-form} \,, \\  
  R\,\llcorner & :& \mbox{$p$-form} &\to& \mbox{$(p-3)$-form} \,.
  \end{array}
}
These operators can be conveniently summarized using a generalized derivative 
 of the form \cite{Shelton:2005cf}
\eq{
  \mathcal D = d + H \wedge + F\circ + Q\bullet + R \op\llcorner  \,,
}
where $d$ is the ordinary exterior derivative. 
We furthermore note that $H$, $Q$ and $F_3$ are odd under
the action of the combined world-sheet parity and 
left-moving fermion number $ \Omega_{\rm P}(-1)^{F_{\rm L}} $ whereas
$F$ and $R$ are even  \cite{Shelton:2005cf,Blumenhagen:2015lta}.
When acting with $\mathcal D$ on the cohomology bases shown in \eqref{basis_002} 
we  obtain 
\eq{
  \label{hi_001}
  &\mathcal D\op \bigl(\sigma^0, \omega_{\alpha},\sigma^a\bigr) 
  \equiv \bigl(\mathsf H, \mathsf F_{\alpha},\mathsf Q^a\bigr) \equiv \Xi_A \in H^3_-(\mathcal X) \,,  
  \\[4pt]
  &\mathcal D \op\bigl(\omega_0, \sigma^{\alpha},\omega_a\bigr) 
  \equiv \bigl(\mathsf R, \mathsf Q^{\alpha},\mathsf F_a\bigr) \equiv \Theta^A \in H^3_+(\mathcal X) \,,
}
where we employ the same collective index $A$ as in \eqref{lu_001}.
Note also that the three-forms $F_3$, $\Xi_A$ and $\Theta^A$ can be expanded
in the integral bases \eqref{basis_001},
for which the expansion coefficients are 
quantized due to the familiar flux-quantization condition. 
This includes in particular the quantization of non-geometric fluxes.

%%%%%%%%%%%%%%%%%%%%%%%%%%%%%%%%%%%%%%%%%%%%%%%
%%%%%%%%%%%%%%%%%%%%%%%%%%%%%%%%%%%%%%%%%%%%%%%

\subsubsection*{Bianchi identities and tadpole cancellation condition}

In our setting we assume the absence of localized 
NS-NS sources such as NS5-branes, Kaluza-Klein monopoles or
non-geometric branes, and hence the Bianchi identities for the NS-NS fluxes can be 
expressed as $\mathcal D^2 = 0$ \cite{Shelton:2005cf,Grana:2006hr}. 
Using the three-forms defined in \eqref{hi_001}, 
these conditions can be written as
\eq{
  \label{bianchi_001}
  \int_{\mathcal X} \Xi_A \wedge \Xi_B = 0 \,, 
  \hspace{40pt}
  \int_{\mathcal X} \Theta^A \wedge \Theta^B =0\,,
  \hspace{40pt}
  \Xi_A \otimes \Theta^A = 0\,.
}
Turning to the R-R sector, performing an orientifold projection typically gives rise to orientifold 
planes and requires the introduction of D-branes. Both are 
charged under the R-R potentials and
therefore the Bianchi identities for the R-R field strengths are in general 
non-trivial. Their integrated versions -- or equivalently their formulation in cohomology 
-- are known as the tadpole cancellation 
conditions and read (see for instance \cite{Plauschinn:2018wbo} for more details of this formulation)
\eq{
  \label{hi_002}
  \mathcal D \mathcal F =  \sum_{{\rm D}p+ {\rm D}p'} \mathcal Q_{{\rm D}p}
  + \sum_{{\rm O}p}  \mathcal Q_{{\rm O}p} \,,
}
where the combined R-R field strength is given by $\mathcal F= \mathcal D\op \mathcal C$, $\mathcal Q_{{\rm D}p}$ 
and $\mathcal Q_{{\rm O}p}$ are multi-forms encoding the D-brane and O-plane charges, and  
the sums are over all D-branes, their orientifold images and O-planes present in the background.
We discuss these conditions in more detail in section~\ref{sec_tadpole_charges} below.
In our situation the only non-trivial R-R flux is $F_3$ and 
we obtain for the left-hand side in \eqref{hi_002}
\eq{
  \mathcal D \mathcal F = N^0 \op \omega_0 + N_{\alpha} \op \sigma^{\alpha} + N^a \op\omega_a \,,
}
where 
\eq{
  \label{hi_017}
  N_A = \int_{\mathcal X} \Xi_A \wedge F_3 = \Bigl(  (H\wedge F_3)^0\op,\, (F\circ F_3)_{\alpha}\op,\, (Q\bullet F_3)^a\Bigr)
}
are the contributions of the $H$-, $F$- and $Q$-fluxes to the 
D3-, D5- and D7-brane tadpoles, respectively.  We will refer to the $N_A$ also 
as tadpole charges, and we note that 
since the  $\Xi_A$ and $F_3$ are quantized the $N_A$ take 
integer values.

%%%%%%%%%%%%%%%%%%%%%%%%%%%%%%%%%%%%%%%%%%%%%%%
%%%%%%%%%%%%%%%%%%%%%%%%%%%%%%%%%%%%%%%%%%%%%%%
%%%%%%%%%%%%%%%%%%%%%%%%%%%%%%%%%%%%%%%%%%%%%%%
%%%%%%%%%%%%%%%%%%%%%%%%%%%%%%%%%%%%%%%%%%%%%%%

\subsection{Scalar potential}

The fluxes introduced above generate a scalar potential $V=V_{\rm F}+V_{\rm D}$ in the effective four-di\-men\-sio\-nal 
$\mathcal N=1$ supergravity theory. 
The F-term potential $V_{\rm F}$ is expressed in terms of a superpotential $W$ 
and the K\"ahler potential \eqref{lu_005}
as
\eq{
  \label{pot_003}
  V_{\rm F} = e^K \left[  D_{i\vphantom{\ov j}} W\op G^{i\ov j} D_{\ov j} \ov W
  + D_{A\vphantom{\ov B}} W\op G^{A\ov B} D_{\ov B} \ov W - 3 |W|^2 \right],
}
where $D_iW$ and  $D_AW$ are the K\"ahler-covariant derivatives 
and $G^{i\ov j}$ and $G^{A\ov B}$ denote the inverse K\"ahler metrics in 
the complex-structure and K\"ahler sector, respectively. 
The fluxes generate 
a  generalization of the Gukov-Vafa-Witten superpotential of the form
$W = \int_{\mathcal X} \Omega \wedge (  F_3 - \mathcal D\op\Phi )$
\cite{Shelton:2005cf,Villadoro:2006ia,Shelton:2006fd}
which, using \eqref{hi_001} and \eqref{hi_007}, can be expressed as
\eq{
\label{pot_002}
 W = \int_{\mathcal X} \Omega\wedge G\,,
 \hspace{50pt}
  G =    
  F_3 - \Xi_A\op \mathsf T^A \,.
}
The D-term potential in the effective theory takes the  form
$V_{\rm D} = [ \mbox{Re}\op f \op]^{-1|\kappa\lambda} D_{\kappa} D_{\lambda}$,
where $f_{\kappa\lambda}$ is the gauge-kinetic function for the $U(1)$ vector fields 
shown in table~\ref{lu_007}. We emphasize that $f_{\kappa\lambda}$ only depends on the 
complex-structure moduli, and we note that 
the D-term potential is generated by the fluxes 
$\Theta^A$ \cite{Robbins:2007yv,Shukla:2015bca,Blumenhagen:2015lta}.
It can be written as 
\eq{
  \label{pot_001}
  V_{\rm D} = 2 \int_{\mathcal X} \bigl( \widetilde\Theta^A K_{A\vphantom{\ov B}}  \bigr)\wedge\star 
  \bigl( \widetilde\Theta^{\ov B}   K_{\ov B} \bigr)
  \,,
}
where, due to the self-duality of the R-R four-form $C_4$ (see \cite{Grimm:2004uq} for details),  
only the restricted three-forms $\widetilde\Theta^A= - \Theta^A{}_{\lambda} \op\beta^{\lambda}$ 
appear.

%%%%%%%%%%%%%%%%%%%%%%%%%%%%%%%%%%%%%%%%%%%%%%%
%%%%%%%%%%%%%%%%%%%%%%%%%%%%%%%%%%%%%%%%%%%%%%%
%%%%%%%%%%%%%%%%%%%%%%%%%%%%%%%%%%%%%%%%%%%%%%%
%%%%%%%%%%%%%%%%%%%%%%%%%%%%%%%%%%%%%%%%%%%%%%%
%%%%%%%%%%%%%%%%%%%%%%%%%%%%%%%%%%%%%%%%%%%%%%%
%%%%%%%%%%%%%%%%%%%%%%%%%%%%%%%%%%%%%%%%%%%%%%%

\section{Tadpole charges}
\label{sec_tadpole_charges}

In equation \eqref{hi_017}  we have defined 
tadpole charges $N_A$, which  play an important role 
for moduli stabilization. They connect the open- and closed-string sectors
to each other and give rise  to strong restrictions 
(see for instance~\cite{
Betzler:2019kon,
Bena:2019sxm,
Carta:2020ohw,
Demirtas:2020ffz,
Blumenhagen:2020ire,
Gao:2020xqh,
Braun:2020jrx,
Bena:2020xrh,
Crino:2020qwk
}
for recent discussions of this question). 
In this section we present arguments that the  $N_A$
should contribute to the 
tadpole cancellation conditions
\eqref{hi_002}
in the same way as D-branes and not as anti-D-branes. This implies
\eq{
  N_A \leq 0  \,.
}

%%%%%%%%%%%%%%%%%%%%%%%%%%%%%%%%%%%%%%%%%%%%%%%
%%%%%%%%%%%%%%%%%%%%%%%%%%%%%%%%%%%%%%%%%%%%%%%

\subsubsection*{D-branes and orientifold planes}

Let us start by extending our review of type IIB orientifold 
compactifications from section~\ref{sec_prelim_set}.
When performing an orientifold projection 
of the form  $(-1)^{F_{\rm L}} \Omega_{\rm P}\op\sigma$ 
the fixed loci of $\sigma$  correspond to orientifold planes
which  fill-out four-dimensional space-time $\mathbb R^{3,1}$ and 
wrap cycles $\Gamma_{{\rm O}p}$ in the compact space $\mathcal X$.
In our case these are O3- and O7-planes which are point-like and 
are wrapping four-cycles in $\mathcal X$, respectively. 
When orientifold planes are present then D-branes can be introduced,
which in our situation are  D3- and D7-branes
filling-out four-dimensional space-time $\mathbb R^{3,1}$ 
and wrapping cycles $\Gamma_{{\rm D}p}$ in $\mathcal X$.
Additionally, D-branes can have a non-trivial gauge flux $\mathcal F_{{\rm D}p}
=  \mathsf F_{{\rm D}p} + B \in H^2(\Gamma_{{\rm D}p})$ 
on their world-volume, where $\mathsf F_{{\rm D}p}$ denotes the  
open-string gauge flux  and 
$B$ is understood as pulled-back to the D-brane. We do not consider discrete torsion for the $B$-field.

The data characterizing D-branes and O-planes can be encoded in the charges $\mathcal Q_{{\rm D}p}$ and 
$\mathcal Q_{{\rm O}p}$ which already appeared in \eqref{hi_002}. These are multi-forms defined as
\eq{
  \mathcal Q_{{\rm D}p} &=  \lambda \left[     [\Gamma_{{\rm D}p}] \wedge
   \mbox{tr} \left( e^{\mathsf F_{{\rm D}p}} \right)  \wedge
    \sqrt{ \frac{\hat{\mathcal A}(\mathcal R_T)}{\hat{\mathcal A}(\mathcal R_N)}}
   \right] ,
  \\
  \mathcal Q_{{\rm O}p} &
  = Q_p \,\lambda\left[  [\Gamma_{{\rm O}p}]\wedge \sqrt{ \frac{\mathcal{L}(\mathcal R_T/4)}{\mathcal{L}(\mathcal R_N/4)}} 
   \right] ,
}
where $[\Gamma_{{\rm D}p}]$ is the Poincar\'e dual of the cycle wrapped by the D-brane,
$\mathsf F_{{\rm D}p}$ denotes the quantized open-string gauge-flux  and the trace is over the fundamental representation.
The expressions $\mathcal R_T$ and $\mathcal R_N$ stand for the restrictions of the 
curvature two-form $\mathcal R$ to the 
tangent and normal bundle of $\Gamma_{{\rm D}p}$, and we used
the $\hat{\mathcal A}$-genus and the Hirzebruch polynomial $\mathcal{L}$.
The charge $Q_p$ of the orientifold planes is given by
$Q_p=-2^{p-4}$, and 
the operator $\lambda$ was introduced below equation \eqref{hi_071}.

%%%%%%%%%%%%%%%%%%%%%%%%%%%%%%%%%%%%%%%%%%%%%%%
%%%%%%%%%%%%%%%%%%%%%%%%%%%%%%%%%%%%%%%%%%%%%%%

\subsubsection*{Calibrations}

In order for D-branes to preserve $\mathcal N=1$ supersymmetry,  
calibration conditions have to be satisfied. In our situation they are
of the form \cite{Marino:1999af} (see for instance  \cite{Jockers:2005pn} for a review)\footnote{
It would be desirable to include curvature terms in the calibration condition, but
we are not aware of any work addressing this question.}
\eq{
  \label{calib_002}
  {\rm Vol}(\Gamma_{{\rm D}p}) = e^{i\op \theta_{{\rm D}p}} \int_{\Gamma_{{\rm D}p}} 
  e^{-\phi}\,\lambda \left[ \op
  \mbox{tr}\bigl(e^{\mathcal F_{{\rm D}p}} \bigr) 
  \wedge  e^{- i\op J} \right] 
  ,
}
where the volume is computed using the DBI action and $J$ denotes the K\"ahler form of $\mathcal X$.
The operator $\lambda$ was introduced below equation \eqref{hi_071},
and $\theta_{{\rm D}p}$ is an up to now undetermined  phase which selects a particular $\mathcal N=1$ combination 
of $\mathcal N=2$ supercharges. 
Turning to orientifold planes, since the involution $\sigma$ is holomorphic the cycles $\Gamma_{{\rm O}p}$
are holomorphic and  satisfy the calibration conditions 
\eq{
  \label{calib_001}
  {\rm Vol}(\Gamma_{{\rm O}p}) = e^{i\op \theta_{{\rm O}p}} \int_{\Gamma_{{\rm O}p}} 
  e^{-\phi}\,\lambda \left[ 
  e^{- i\op J} \right]
  ,
  \hspace{60pt}
  \theta_{{\rm O}p} = 0\,.
}
Note that since a gauge flux $\mathcal F_{{\rm D}p}$ is odd under the orientifold projection $\sigma$ it vanishes 
when restricted to the fixed-point set of $\sigma$. We can therefore trivially extend
\eqref{calib_001} by $\mbox{tr}\op(e^{\mathcal F_{{\rm D}p}})$ and compare with \eqref{calib_002}.
In order for all D-branes and O-planes to preserve the same $\mathcal N=1$ supersymmetry,
we then have to require
\eq{
  \theta_{{\rm D}p} = \theta_{{\rm O}p} =0  \,.
}

%%%%%%%%%%%%%%%%%%%%%%%%%%%%%%%%%%%%%%%%%%%%%%%
%%%%%%%%%%%%%%%%%%%%%%%%%%%%%%%%%%%%%%%%%%%%%%%

\subsubsection*{Tadpole cancellation conditions}

The tadpole cancellation conditions  have been stated schematically already in \eqref{hi_002},
but here we want to make them more precise.
For our setting with \mbox{O3-/O7-}planes and D3-/D7-branes the tadpole cancellation 
conditions  can be written in the following way 
(see for instance  \cite{Plauschinn:2008yd} for details of the derivation)
\begin{subequations}
\label{tadpole_all}
\begin{align}
\label{tadpole_d7}
&N^a \omega_a = \sum_{{\rm D}7_{\mathsf i}} N_{{\rm D}7_{\mathsf i}}
  \Bigl(\op  [\Gamma_{{\rm D}7_{\mathsf i}}] +[\Gamma'_{{\rm D}7_{\mathsf i}}]\Bigr) - 8\op \sum_{{\rm O}7_{\mathsf j}}\: 
  [\Gamma_{{\rm O}7_{\mathsf j}}]\,,
\\[4pt]
\label{tadpole_d5}
&N_{\alpha} \sigma^{\alpha} = -\sum_{{\rm D}7_{\mathsf i}} \Bigl(
    \mbox{tr}\left[  \mathsf F_{{\rm D}7_{\mathsf i}}\right] \wedge
    \left[ \Gamma_{{\rm D}7_{\mathsf i}} \right]
    +
    \mbox{tr}\left[  \mathsf F'_{{\rm D}7_{\mathsf i}}\right] \wedge
    \left[ \Gamma'_{{\rm D}7_{\mathsf i}} \right]
    \Bigr)\,,
\\    
\label{tadpole_d3}
&\frac{N^0}{2}  =  N_{{\rm D}3}  -  \frac{N_{{\rm O}3}}{4} 
  -\sum_{{\rm D}7_{\mathsf i}}\left( \frac{1}{2}\int_{\Gamma_{{\rm D}7_{\mathsf i}}}
  \hspace{-10pt}
    \mbox{tr}\left[ \mathsf F^2_{{\rm D}7_{\mathsf i}}\right] 
  + N_{{\rm D}7_{\mathsf i}}\op\frac{\chi( \Gamma_{{\rm D}7_{\mathsf i}}\bigr)}{24} 
  \right)
  -\sum_{{\rm O}7_{\mathsf j}} \frac{\chi\bigl( \Gamma_{{\rm O}7_{\mathsf j}}\bigr)}{12} \,,
\end{align}
\end{subequations}
where $N_{{\rm D}7_{\mathsf i}}$ denotes the number of D7-branes in a stack labelled by $\mathsf i$
and $N_{{\rm D}3}$ is the total number of D3-branes. Both of these numbers are counted 
without the orientifold images which are denoted by a prime.
Furthermore, $N_{{\rm O}3}$ is the total number of O3-planes, and $\chi(\Gamma)$ denotes 
the Euler number of the cycle $\Gamma$.\footnote{
In case the D7-branes have double intersection points
the corresponding Euler number has to be  corrected by the number of pinch points
\cite{Aluffi:2007sx,Collinucci:2008pf}. This subtlety will however not be relevant here.}

%%%%%%%%%%%%%%%%%%%%%%%%%%%%%%%%%%%%%%%%%%%%%%%
%%%%%%%%%%%%%%%%%%%%%%%%%%%%%%%%%%%%%%%%%%%%%%%

\subsubsection*{Gauge groups}

Let us also recall that  D-branes give rise to gauge theories on their world-volume.  In 
type IIB the corresponding gauge groups are either $U(N_{{\rm D}p})$, $SO(2N_{{\rm D}p})$ or $Sp(2N_{{\rm D}p})$,
where $N_{{\rm D}p}$ denotes the number of D$p$-branes on top of each other. 
Since we can have several such stacks of D-branes wrapping cycles in the compact
space, the four-dimensional gauge group is of the schematic form
\eq{
  \label{tadpole_g}
  \mathsf G = \prod_{\mathsf i} U(N_{{\rm D}p_{\mathsf i}}) \times 
  \prod_{\mathsf j} SO(2N_{{\rm D}p_{\mathsf j}}) \times 
  \prod_{\mathsf k} Sp(2N_{{\rm D}p_{\mathsf k}}) \times U(1)^{h^{2,1}_+} \,,
}
where the last factor corresponds to the closed-string $U(1)$ gauge fields $U^{\lambda}$ 
mentioned in table~\ref{lu_007}.

Now, in \cite{Vafa:2005ui} conjectures have been presented which imply  
that the rank of the total gauge group in a consistent quantum-gravity theory should be 
bounded from above. 
This requirement is supported for instance by extensive experience in string-theory model building, and 
progress in establishing 
such a bound for different settings 
has been made for instance in \cite{Park:2011wv,Kim:2019vuc,Lee:2019skh}.
And indeed, in the absence of fluxes we see from \eqref{tadpole_all}
that the rank of the gauge group $\mathsf G$ 
is  bounded by the orientifold data.
More concretely, for  a fixed compactification space and orientifold projection, 
for supersymmetric D-brane configurations and  in the absence of closed-string fluxes, 
the rank of $\mathsf G$ 
is bounded by the number of O-planes, 
by the Euler numbers of four-cycles in $\mathcal X$ and 
by $h^{2,1}_+$.
Our main requirement is then that also when introducing  fluxes,
the rank of the gauge group $\mathsf G$ remains bounded from above.

%%%%%%%%%%%%%%%%%%%%%%%%%%%%%%%%%%%%%%%%%%%%%%%
%%%%%%%%%%%%%%%%%%%%%%%%%%%%%%%%%%%%%%%%%%%%%%%

\subsubsection*{Contribution of closed-string fluxes}

We now want to argue that the closed-string fluxes encoded in the tadpole charges $N_A$ 
should contribute in the same way as supersymmetric D-branes to the tadpole cancellation conditions \eqref{tadpole_all}
and not as  anti-D-branes or orientifold planes. 
Loosely speaking,  in string theory one should not be able to generate arbitrarily-large gauge groups. 
\begin{itemize}

\item Let us start by considering the D3-brane tadpole condition 
and assume that we found a configuration of branes and fluxes which solves \eqref{tadpole_d3},
say for $N^0=0$.
A new solution can be obtained by changing the fluxes and number of D3-branes as\footnote{
As a physical process this is known as brane-flux transmutation \cite{Kachru:2002gs},
however, in general \eqref{tadpole_830} will lead to a physically-different configuration solving
the tadpole condition \eqref{tadpole_d3}.}
\eq{
  \label{tadpole_830}
  N^0\to N^0 + 2\op\alpha\,, \hspace{30pt}
  N_{{\rm D}3}\to N_{{\rm D}3} +\alpha\,, \hspace{60pt}
  \alpha\in\mathbb Z\,.
}
Since for supersymmetric D-branes $N_{{\rm D3}}$ is  non-negative (cf.~the calibration condition 
\eqref{calib_002}), the integer $\alpha$ is bounded from below.  
Requiring furthermore the rank of the gauge group to be bounded implies that $\alpha$ also has to be bounded 
from above. These restrictions then translate into bounds on $N^0$.
It is not clear to us how to derive an upper bound on $N^0$ from first principles, but since the 
tadpole cancellation conditions contain only topological data in the form of integers,
the only natural bound is
\eq{
  \label{charge_01}
  N^0 \leq 0 \,.
}
We explain this point in more detail below. 
This means that the combination of fluxes contained in $N^0$ contributes like a D-brane to the 
tadpole conditions and not like an anti-D-brane or orientifold plane. 
Note also that  \eqref{charge_01} is of course realized in standard flux compactifications.

\item A similar reasoning can  be applied to the D7-brane tadpole 
\eqref{tadpole_d7}. For simplicity we consider a setting
with $h^{1,1}_-=0$ and $\chi( \Gamma_{{\rm D}7})=0$,
which is satisfied for instance in toroidal-orbifold compactifications. 
Let us then again assume a configuration of D$7$-branes
has been found which solves \eqref{tadpole_d7} for $N^a = 0$. 
A new solution can be obtained by 
replacing $N^a\to N^a + 2\op\alpha^a$, where  $\alpha^a\in\mathbb Z$,
and introducing/removing  $\alpha^a$ D$7$-branes without gauge flux wrapping 
a cycle dual to  $\omega_{\hat a}$.
The $\alpha^a$ are bounded from below by the requirement that 
$N_{{\rm D}7}\geq 0$, and from above by requiring the rank 
of the gauge group to be bounded. 
It is again not clear to us how to derive an upper bound on the $N^a$,
but a natural choice is 
\eq{
  \label{charge_02}
  N^a \leq 0 \,,
}
and we discuss this point below in more detail.
Let us emphasize that this restriction can be derived from \eqref{charge_01} via T-duality.
From \eqref{charge_02} we then see that the flux combinations $N^a$ contribute like D-branes to the tadpole conditions and not 
like anti-D-branes or orientifold planes. 
Finally, for more general situations with $h^{1,1}_-\neq0$ or $\chi( \Gamma_{{\rm D}7}\bigr)\neq0$ 
also the D3- and 
D5-brane tadpole might have to be adjusted,  but the main 
conclusion doesn't change.

\item For the D5-brane tadpole \eqref{tadpole_d5} we have to 
slightly extend our setting  by including
D5-branes. In this case the corresponding tadpole condition has been determined for instance
in \cite{Plauschinn:2008yd}, and following similar arguments as above we can infer that 
\eq{
    N_{\alpha} \leq 0 \,.
}
\end{itemize}
To summarize, we have argued that in order for the total rank of the gauge group (shown in
\eqref{tadpole_g}) to be bounded -- in agreement with the swampland conjectures made in \cite{Vafa:2005ui} -- 
the tadpole charges $N_A$ should be bounded from 
above. 
We  identified 
\eq{
  \label{tadpole_831}
  \bom{
   N_A\leq 0
   }
}
as the most natural choice for such a bound, which is consistent 
with known constructions as well as with T-duality. This
implies that fluxes should contribute in the same way as supersymmetric 
D-branes to the tadpole cancellation conditions.

%%%%%%%%%%%%%%%%%%%%%%%%%%%%%%%%%%%%%%%%%%%%%%%
%%%%%%%%%%%%%%%%%%%%%%%%%%%%%%%%%%%%%%%%%%%%%%%

\subsubsection*{Identifying a natural bound for integers}

We mentioned above that for integers  a natural threshold  is given by zero, which we now would like to 
explain in more detail. 
To do so, let us first consider a setting with a dimension-full quantity such as an energy $E$ which is bounded 
from above. If for this setting 
there exists a single characteristic energy scale $E_0$, then one would naturally choose $E_0$ as the 
upper bound for $E$. That is $E\leq E_0$, or equivalently $E/E_0\leq 1$.
Similarly, let us consider a setting with a positive dimension-less quantity $\epsilon\in\mathbb R^+$
which is known to be bounded from above.
In the absence of any related characteristic quantity (such as $2\pi$), a natural upper bound for $\epsilon$ would be
given by 
$\epsilon \leq 1$. This would, for instance, typically allow for a power-series expansion.
Finally, we consider a setting with a dimension-less integer  $p\in\mathbb Z$ \textit{without} a related 
characteristic quantity. 
This situation is different from  the previous two cases, since here we would not 
be able to argue via a natural scale in the problem or 
via a power-series expansion. 
However, the only special point is $p_0=0$ where the sign of $p$ changes, and 
we can identify this point as a natural threshold. 
Hence, when forced to choose an upper bound for integers (taking positive as well as negative values) 
in a setting without any characteristic quantity, we identify $p\leq 0$ is the most natural choice.

%%%%%%%%%%%%%%%%%%%%%%%%%%%%%%%%%%%%%%%%%%%%%%%
%%%%%%%%%%%%%%%%%%%%%%%%%%%%%%%%%%%%%%%%%%%%%%%

\subsubsection*{Discussion}

As we have emphasized above, with our current knowledge we are not able to 
derive the requirement $N_A\leq 0$ on the tadpole charges from first principles. 
This condition has therefore the character of a (swampland) conjecture,
which is however based on plausability and duality arguments and on requiring 
consistency with other swampland conjectures.

Naturally, if $N_A\leq0$ is a non-empty statement then some constructions which have appeared 
in the literature will not satisfy this condition. Let us discuss a selection of them:
\begin{itemize}

\item In some of the examples in \cite{Camara:2005dc} and \cite{Aldazabal:2006up} 
closed-string fluxes contribute as anti-D-branes to the tadpole cancellation 
condition, and hence  violate our requirement $N_A\leq 0$.
In these papers the possibility to have fluxes contributing as anti-D-branes is considered 
to be an advantage which allows for more flexibility in model building, but
we take the opposite point of view. However, with our current knowledge we are
not able to exclude these models from first principles.

\item Some of the AdS vacua found in  \cite{Font:2008vd} and \cite{Dibitetto:2011gm}
do not satisfy the requirement $N_A\leq0$, but some examples 
in these papers  do indeed satisfy this condition.
This shows that one can find non-trivial solutions to the condition $N_A\leq 0$.

\item Similarly,  in \cite{Betzler:2019kon} we studied supersymmetric AdS vacua for the 
$\mathbb T^6/\mathbb Z_2\times \mathbb Z_2$ type IIB orientifold and 
found solutions with $N_A\leq 0$ and all moduli stabilized (see equation~(5.14)
in  \cite{Betzler:2019kon} and note that our conventions are related as $N_{\rm here} = - \mathsf Q_{\rm there}$).
For this paper we randomly generated $3\cdot 10^6$ stable AdS vacua with all moduli stabilized,
but of those only a fraction of $1.57\cdot 10^{-4}$ satisfy 
the requirement $N_A\leq 0$. This is a considerable reduction of this part of the string-theory landscape.

\end{itemize}

%%%%%%%%%%%%%%%%%%%%%%%%%%%%%%%%%%%%%%%%%%%%%%%
%%%%%%%%%%%%%%%%%%%%%%%%%%%%%%%%%%%%%%%%%%%%%%%

\subsubsection*{Remarks}

We close this section with the following comments and remarks concerning the 
tadpole charges $N_A$:
\begin{itemize}

\item Let us recall the general form of the tadpole cancellation conditions 
\eqref{hi_002}, multiply both sides with
$e^{-\phi}\op\lambda \op [   e^{B - i\op J} ]$ and integrate over $\mathcal X$.
Using the definition of the moduli shown in 
\eqref{hi_007} and the calibration conditions \eqref{calib_002} and 
\eqref{calib_001}, we  obtain
\eq{
  \frac{N_A \op \tau^A}{2} =&
  \arraycolsep2pt
  \begin{array}[t]{cl@{\op}lclr@{\op}lcll}
  &\displaystyle e^{-\phi} & \displaystyle N_{{\rm D}3}  &+& 
  \displaystyle \sum_{{\rm D}7_{\mathsf i}} \,\biggl[ & &
  \displaystyle \mbox{Vol}(\Gamma_{{\rm D}7_{\mathsf i}}) 
  &- &
  \displaystyle e^{-\phi} \, N_{{\rm D}7_{\mathsf i}} \,\frac{\chi(\Gamma_{{\rm D}7_{\mathsf i}})}{24}
  & \displaystyle\biggr]
  \\
  - &\displaystyle  e^{-\phi} & \displaystyle \frac{N_{{\rm O}3}}{4}
  &-& \displaystyle \sum_{{\rm O}7_{\mathsf j}} \,\biggl[ &
  4 & \displaystyle  \op\mbox{Vol}(\Gamma_{{\rm O}7_{\mathsf j}}) 
  &+& \displaystyle e^{-\phi} \, \frac{\chi(\Gamma_{{\rm O}7_{\mathsf j}})}{12}
  & \displaystyle\biggr] \,.
  \end{array} 
}
Our requirement of having the closed-string fluxes contributing in the same way as 
supersymmetric D-branes
to the tadpole conditions (the first line on the right-hand side) then implies
\eq{
  \label{calib_007}
  N_A\tau^A \leq 0 \,.
}

\item In the absence of non-geometric fluxes, the presence of O7-planes requires the 
introduction of D7-branes via the tadpole cancellation condition. However, 
with non-vanishing $Q$-flux we can obtain configurations without D-branes. 
In particular, in the absence of D-branes the tadpole conditions \eqref{tadpole_all} read
\eq{
  N^a \omega_a =  - 8\op \sum_{{\rm O}7_{\mathsf j}}\,  [\Gamma_{{\rm O}7_{\mathsf j}}]\op,
  \hspace{22.25pt}
  N_{\alpha} = 0 \op,
  \hspace{22.25pt}
N^0 =    -  \frac{N_{{\rm O}3}}{2} 
  -\sum_{{\rm O}7_{\mathsf j}} \frac{\chi\bigl( \Gamma_{{\rm O}7_{\mathsf j}}\bigr)}{6} \op,
}
which satisfy the requirement $N_A\leq 0$. 
Note that especially with respect to stabilizing all moduli, this is a desirable situation since
no open-string moduli (typically coming with the introduction of D-branes) need to be stabilized.

\item In this section we have considered effective quantum-gravity theories 
which arise from compactifying string theory in the presence of fluxes and space-time filling D-branes. 
We argued that in this situation the rank of the gauge group should be bounded, 
however, in other situations this does not need to be true. 
For instance, the standard example for the AdS/CFT correspondence
is given by taking $N_{{\rm D}3}$ coincident D3-branes in flat ten-dimensional 
space-time and making $N_{{\rm D}3}$ large. Here there is no restriction on $N_{{\rm D}3}$, though
the resulting theory is dual to a non-gravitational theory.\footnote{We thank 
D.~Junghans for bringing this question to our attention.}

\end{itemize}

%%%%%%%%%%%%%%%%%%%%%%%%%%%%%%%%%%%%%%%%%%%%%%%
%%%%%%%%%%%%%%%%%%%%%%%%%%%%%%%%%%%%%%%%%%%%%%%
%%%%%%%%%%%%%%%%%%%%%%%%%%%%%%%%%%%%%%%%%%%%%%%
%%%%%%%%%%%%%%%%%%%%%%%%%%%%%%%%%%%%%%%%%%%%%%%
%%%%%%%%%%%%%%%%%%%%%%%%%%%%%%%%%%%%%%%%%%%%%%%
%%%%%%%%%%%%%%%%%%%%%%%%%%%%%%%%%%%%%%%%%%%%%%%

\section{Moduli stabilization}
\label{sec_mod_stab}

In this section we discuss some general features of moduli stabilization with 
non-geometric fluxes. This question has been studied before,
for  instance in
\cite{
Shelton:2005cf,
Shelton:2006fd,
Ihl:2007ah,
deCarlos:2009fq,
deCarlos:2009qm,
Danielsson:2012by,
Blaback:2013ht,
Blaback:2015zra
}
for toroidal orbifold compactifications and 
in 
\cite{
Micu:2007rd,
Blumenhagen:2015kja,
Blumenhagen:2015jva,
Andriot:2016ufg,
Shukla:2019wfo,
Marchesano:2020uqz
} 
for more general settings. 
Our results in this section are the following:
in subsection~\ref{sec_nods} 
we determine  for Calabi-Yau orientifold compactifications
the form of the scalar potential
at an extremum as
\eq{
  V\op\Bigr\rvert_{\rm ext} 
  = -\frac{e^{K_{\rm K}}}{2} \left[  \int_{\mathcal X} \bigl( \mbox{Re}\, G \wedge\star  \mbox{Re}\, 
  G \bigr)+ 
  N_A \op\tau^A \right]_{\rm ext},
}
where $G$ denotes the three-form flux defined in \eqref{pot_002}, $\tau^A=\mbox{Im}\,\mathsf T^A$ and 
$N_A$ denotes the tadpole charges. This surprisingly simple form will be employed 
later on. 
In subsection~\ref{sec_hodge_numbers} we show that in order 
to stabilize all moduli by non-geometric fluxes, the Hodge numbers
have to satisfy
\eq{
  h^{1,1} \leq h^{2,1}_- \,, \hspace{60pt} h^{2,1}_+ = 0 \,.
}

%%%%%%%%%%%%%%%%%%%%%%%%%%%%%%%%%%%%%%%%%%%%%%%
%%%%%%%%%%%%%%%%%%%%%%%%%%%%%%%%%%%%%%%%%%%%%%%
%%%%%%%%%%%%%%%%%%%%%%%%%%%%%%%%%%%%%%%%%%%%%%%
%%%%%%%%%%%%%%%%%%%%%%%%%%%%%%%%%%%%%%%%%%%%%%%

\subsection{Non-geometric flux vacua}
\label{sec_nods}

In this section we discuss necessary conditions for (non-)geometric flux vacua
for the setting introduced in section~\ref{sec_prelim}.
We first determine an expression  for extrema of the scalar potential and 
derive a necessary relation for stable minima.

%%%%%%%%%%%%%%%%%%%%%%%%%%%%%%%%%%%%%%%%%%%%%%%
%%%%%%%%%%%%%%%%%%%%%%%%%%%%%%%%%%%%%%%%%%%%%%%

\subsubsection*{Rewriting the F-term  potential}

The scalar potential in $\mathcal N=1$ supergravity is given by $V= V_{\rm F} + V_{\rm D}$,
where the D-term potential has been given in \eqref{pot_001}. 
The F-term potential \eqref{pot_003} for the superpotential \eqref{pot_002}
can be worked out using the relation \eqref{lu_009} as
\eq{
\label{hi_012}
 V_{\rm F} = e^{K} \left[ 
  \,\frac{e^{-K_{\rm cs}}}{2}  \int_{\mathcal X} G\wedge \bigl( \star \ov G + i\op \ov G\bigr)
  +  e^{-2K_{\rm cs}}\, \Xi^{\ov 0}_{A\vphantom{\ov B}} \, \omega^{A\ov B}\,\Xi^0_{\ov B}
  \right],
}
where the three-form flux reads $G =   F_3 - \mathsf T^A \op\Xi_A$ and where
$\Xi^{\ov 0}_{A\vphantom{\ov B}}$ and $\Xi^0_{\ov B}$ denote the $\ov\Omega$ and $\Omega$  components 
of $\Xi_{A\vphantom{\ov B}}$ and $\Xi_{\ov B}$ as in \eqref{lu_010}.
These components together with the matrix $\omega^{A\ov B}$ are given by
\eq{
  \label{hi_014}
    \Xi^{\ov 0}_{A\vphantom{\ov a}}= i\op e^{K_{\rm cs}} \int_{\mathcal X} \Omega \wedge \Xi_A 
  \,, \hspace{60pt}
  \omega^{A\ov B} =  G^{A\ov B} - K^A K^{\ov B} \,,
}
where the former only depends on the complex-structure 
moduli and the latter  only depends on the imaginary parts $\tau^A$ of the K\"ahler-sector moduli. 
(In equation \eqref{omeg_ex} we show the form of $\omega^{A\ov B}$ for a particular example.)
In order to make \eqref{hi_012} more symmetric we can use  $\star^2=-1$ when acting on 
three-forms on an Euclidean six-dimensional space and rewrite the scalar potential as
\eq{
\label{hi_012b}
 V_{\rm F} = e^{K} \left[ 
  \,\frac{e^{-K_{\rm cs}}}{4}  \int_{\mathcal X} (G+ i \star G) \wedge \star \bigl( \ov G - i\star \ov G\bigr)
  +  e^{-2K_{\rm cs}}\, \Xi^{\ov 0}_{A} \, \omega^{A\ov B}\,\Xi^0_{\ov B}
  \right].
}
When ignoring $\alpha'$-corrections to the K\"ahler potential
and  only considering $H$-flux, the second term in  \eqref{hi_012b} vanishes.
This corresponds to the situation  studied  by Giddings, Kachru and Polchinski
in \cite{Giddings:2001yu}.

%%%%%%%%%%%%%%%%%%%%%%%%%%%%%%%%%%%%%%%%%%%%%%%
%%%%%%%%%%%%%%%%%%%%%%%%%%%%%%%%%%%%%%%%%%%%%%%

\subsubsection*{Some identities}

In order to proceed, we determine some identities 
for derivatives of the K\"ahler potential and the K\"ahler metric in the K\"ahler-moduli sector.
We first note that the expression $K_A$ and $K^A$ 
are purely imaginary and hence $K_{A\vphantom{\ov A}} = - K_{\ov A}$
and $K^{A\vphantom{\ov A}} = - K^{\ov A}$.
Using then the relations shown in equation  \eqref{lu_006}, we  compute
\eq{
  \label{lu_015}
  &K^C \partial_C K^A = K^C (-\delta_C{}^A) = - K^A \,,
  \\
  &K^C \partial_C K^{\ov A} = K^C (+\delta_C{}^{\ov A}) = - K^{\ov A} \,,
}
and in a similar way we determine for the first derivatives of the K\"ahler potential
\eq{
  \label{lu_016}
  K^C \partial_C K_A = - K^C G_{C\ov A}  =  +K_A\,,
  \\
  K^C \partial_C K_{\ov A} = +  K^C G_{C\ov A}  =  +K_{\ov A}\,.
}
Furthermore, for the K\"ahler metric $G_{A\ov B}$ and 
for the matrix $\omega^{A\ov B}$ 
defined in \eqref{hi_014}
we compute
\eq{
  \label{lu_017}
  \arraycolsep2pt
  \begin{array}{lclcl}
  \displaystyle K^C \partial_C\op  G_{A\ov B} &=&\displaystyle   K^C \partial_A G_{C\ov B} =
  \partial_A (K^C  G_{C\ov B} ) - (\partial_A K^C) G_{C\ov B} &=& +2\op G_{A\ov B} \,,
  \\[4pt]
  \displaystyle  K^C \partial_C \op \omega^{A\ov B} &=& \displaystyle 
  - G^{A\ov M} (K^C \partial_C G_{\ov M N}) G^{N\ov B} + 2 K^{A\vphantom{\ov B}} K^{\ov B} &=&
  - 2\op \omega^{A\ov B}\,.
  \end{array}
}

%%%%%%%%%%%%%%%%%%%%%%%%%%%%%%%%%%%%%%%%%%%%%%%
%%%%%%%%%%%%%%%%%%%%%%%%%%%%%%%%%%%%%%%%%%%%%%%

\subsubsection*{A necessary condition for extrema}

In order to obtain extrema of the combined potential $V = V_{\rm F} + V_{\rm D}$ 
we have to solve $ \partial_iV=0$ and $ \partial_A V=0$,
where $i$ and $A$ label the complex-structure and combined K\"ahler moduli. 
These are in general 
complicated equations which are typically solved numerically, however, let us 
consider the necessary condition 
\eq{
  0 = K^A \partial_A V  \,.
}
With the help of the relations \eqref{lu_015}, \eqref{lu_016} and \eqref{lu_017} 
we find that the D-term potential satisfies
$  K^A \partial_A V_{\rm D} = 2\op V_{\rm D}$, and for the full potential we obtain
\eq{
  \label{lu_011}
  0 = K^A \partial_A V = 2\op V+\frac{e^{K_{\rm K}}}{2}
  \int_{\mathcal X} (G+ \ov G)\wedge \bigl( \star \ov G + i\op \ov G\bigr) \,.
}
Note that the last term in \eqref{lu_011}
contains an imaginary contribution which, since the scalar potential $V$ is real, implies
\eq{
  \label{lu_018}
 0 = \int_{\mathcal X} (G+\ov G) \wedge \star (G- \ov G) \,.
} 
We now use this relation together with the Bianchi identity \eqref{bianchi_001} 
and the definition of $N_A$ and $\tau^A$ given in \eqref{hi_017} and \eqref{lu_001}
to evaluate the scalar potential at the extremum as
\eq{
  \label{hi_011}
  \bom{
  V\op\Bigr\rvert_{\rm extr} = -\frac{e^{K_{\rm K}}}{2} \left[  \int_{\mathcal X} \bigl( \mbox{Re}\, G \wedge\star  \mbox{Re}\, 
  G \bigr)+ 
  N_A \op\tau^A \right]_{\rm extr}\,.
  }
}

%%%%%%%%%%%%%%%%%%%%%%%%%%%%%%%%%%%%%%%%%%%%%%%
%%%%%%%%%%%%%%%%%%%%%%%%%%%%%%%%%%%%%%%%%%%%%%%

\subsubsection*{A necessary condition for stability}

Next, we derive a necessary condition for stability of the extremum. The combined mass matrix for the K\"ahler-sector moduli $\mathsf T^A$ and complex-structure moduli
$z^i$ takes the following general form
\eq{
  \label{hi_lu_002}
 M^2 = \left[ 
 \renewcommand{\arraystretch}{1.2}
 \begin{array}{cc|cc}
 m^2_{A\ov B} & m^2_{AB\vphantom{\ov B}} & m^2_{A\ov j\vphantom{\ov B}} & m^2_{Aj\vphantom{\ov B}} \\
 m^2_{\ov A\ov B} & m^2_{\ov AB\vphantom{\ov B}} & m^2_{\ov A\ov j\vphantom{\ov B}} & m^2_{\ov Aj\vphantom{\ov B}} \\ 
 \hline
 m^2_{i \ov B} & m^2_{i B\vphantom{\ov B}} & m^2_{i\ov j\vphantom{\ov B}} & m^2_{ij\vphantom{\ov B}} \\
 m^2_{\ov i \ov B} & m^2_{\ov i B\vphantom{\ov B}} & m^2_{\ov i\ov j\vphantom{\ov B}} & m^2_{\ov ij\vphantom{\ov B}} 
 \end{array}
 \right],
}
where  $m^2_{A\ov B} = \partial_{A\vphantom{\ov B}}\partial_{\ov B} V \rvert_{\rm extr}$ 
and similarly for the other entries.
For our purpose we do not need to determine the complete mass matrix, but 
we are only interested in the expression
\eq{
  \label{lu_019}
  K^{A\vphantom{\ov B}} \op m^2_{A\ov B} \op K^{\ov B} = 
   K^{A\vphantom{\ov B}}K^{\ov B} \partial_{A\vphantom{\ov B}}\partial_{\ov B} V \bigr\rvert_{\rm extr}\,.
}
Using again the relations \eqref{lu_015}, \eqref{lu_016} and \eqref{lu_017}
we compute 
\eq{
   K^{A\vphantom{\ov B}}K^{\ov B} \partial_{A\vphantom{\ov B}}\partial_{\ov B} V
   &=
   K^{A\vphantom{\ov B}} \partial_{A\vphantom{\ov B}} V+ K^{\ov B} \partial_{\ov B} 
   \bigl( K^{A\vphantom{\ov B}} \partial_{A\vphantom{\ov B}} V \bigr) 
   \\ 
   &=
   K^{A\vphantom{\ov B}} \partial_{A\vphantom{\ov B}} V+
   2 K^{\ov B} \partial_{{\ov B}} V
   \\
   &\hspace{20pt}+\frac{e^{K_{\rm K}}}{2}
  \int_{\mathcal X} \Bigl[ \op G\wedge\star G + 5 G\wedge\star \ov G + 2 \ov G \wedge\star \ov G + 3 i \op G\wedge \ov G \op\Bigr]\,,
}
and, employing
\eqref{lu_011} and \eqref{lu_018}, we
evaluate \eqref{lu_019} at the extremum as
\eq{
  \label{hi_lu}
  K^{A\vphantom{\ov B}} \op m^2_{A\ov B} \op K^{\ov B} = 
  \left[ -6\op   V + e^{K_{\rm K}} \int_{\mathcal X} G\wedge \star \ov G \,\right]_{\rm extr}\,.
}
For a stable minimum of the potential  the mass matrix \eqref{hi_lu_002} has to be positive definite, which 
 implies that the left-hand side of \eqref{hi_lu} has to be positive. This allows us to determine the 
 following necessary condition 
\eq{
\label{hi_lu_001}
V\op\bigr\rvert_{\rm min}  <  \frac{e^{K_{\rm K}}}{6} \int_{\mathcal X} G\wedge \star \ov G \;\biggr\rvert_{\rm min} \,.
}

%%%%%%%%%%%%%%%%%%%%%%%%%%%%%%%%%%%%%%%%%%%%%%%
%%%%%%%%%%%%%%%%%%%%%%%%%%%%%%%%%%%%%%%%%%%%%%%

\subsubsection*{Comments on de-Sitter vacua}

Let us briefly return to our discussion of the tadpole charges in section~\ref{sec_tadpole_charges},
where we argued that $N_A \op\tau^A\leq0$.
From the form of the potential in the minimum 
\eqref{hi_011}, we see that a necessary condition for
a de-Sitter minimum is 
\eq{
 0 >  \int_{\mathcal X} \bigl( \mbox{Re}\, G \wedge\star  \mbox{Re}\, 
  G \bigr)+   N_A \op\tau^A \,.
}
Since the Hodge star is positive  definite the first term on the right-hand side is always non-negative,
and hence $N_A\tau^A$ has to be negative for a de-Sitter solution. This shows that our requirement
\eqref{calib_007} of having fluxes contributing like D-branes to the tadpole cancellation conditions
does not exclude  de-Sitter vacua. 
Furthermore, together with the stability condition \eqref{hi_lu_001} we find the following 
constraint at the minimum
\eq{
  0 < 
  - \int_{\mathcal X} \bigl( \mbox{Re}\op G \wedge\star  \mbox{Re}\op  G \bigr) -   N_A \op\tau^A
  < \frac{1}{3}\int_{\mathcal X} \bigl( \mbox{Re}\op G \wedge\star  \mbox{Re}\op G + 
  \mbox{Im}\op G \wedge\star  \mbox{Im}\op  G \bigr).
}
These inequalities restrict stable de-Sitter minima, but they do not exclude them. 
We come back to this relation below. 

%%%%%%%%%%%%%%%%%%%%%%%%%%%%%%%%%%%%%%%%%%%%%%%
%%%%%%%%%%%%%%%%%%%%%%%%%%%%%%%%%%%%%%%%%%%%%%%

\subsubsection*{Remark}

To provide some intuition for the matrix $\omega^{A\ov B}$ defined in \eqref{hi_014}, 
let us consider the type IIB $\mathbb T^6/\mathbb Z_2 \times \mathbb Z_2$ 
orientifold with O3-/O7-planes (where we ignore the twisted sectors). 
This compactification space is characterized 
by $h^{1,1}_+=h^{2,1}_-=3$ and $h^{1,1}_-=h^{2,1}_+=0$, and  the K\"ahler potential for the 
combined K\"ahler-sector
moduli is given by
\eq{
  K=  - \sum_{A=0}^3 \log\left[ - i\op (\mathsf T^A-\ov{\mathsf{T}}{}^A\bigr)\right] .
}
With $\tau^A = \mbox{Im}\op \mathsf T^A$ the matrix $\omega^{A\ov B}$ in \eqref{hi_014} then takes the form
\eq{
  \label{omeg_ex}
  \omega^{A\ov B} = -4 \left[ \begin{array}{cc}
  0 & \tau^A \tau^B \\[2pt] \tau^A \tau^B & 0 \end{array}\right] .
}

%%%%%%%%%%%%%%%%%%%%%%%%%%%%%%%%%%%%%%%%%%%%%%%
%%%%%%%%%%%%%%%%%%%%%%%%%%%%%%%%%%%%%%%%%%%%%%%
%%%%%%%%%%%%%%%%%%%%%%%%%%%%%%%%%%%%%%%%%%%%%%%
%%%%%%%%%%%%%%%%%%%%%%%%%%%%%%%%%%%%%%%%%%%%%%%

\subsection{Constraints on the Hodge numbers}
\label{sec_hodge_numbers}

Let us  now require that all
complex-structure moduli $z^i$ and combined K\"ahler moduli $\mathsf T^A$ 
are stabilized classically by the fluxes introduced in section~\ref{sec_flux_intro}. 
In order to achieve that, Bianchi identities imply  constraints on the 
Hodge numbers which we derive in this section.

%%%%%%%%%%%%%%%%%%%%%%%%%%%%%%%%%%%%%%%%%%%%%%%
%%%%%%%%%%%%%%%%%%%%%%%%%%%%%%%%%%%%%%%%%%%%%%%

\subsubsection*{Notation}

To derive constraints on the Hodge numbers for stabilizing all 
closed-string moduli, it is useful to introduce some more notation. 
First, for ease of presentation, we define the two numbers 
\eq{
  d_1 = 2(h^{2,1}_-+1)\,, \hspace{50pt} 
  d_2 = h^{1,1}+1 \,.
}
Next, for the symplectic basis
$\{\alpha_I,\beta^I\}\in H^3_-(\mathcal X)$ 
shown in \eqref{basis_001} we define two $d_1\times d_1$ matrices
as follows (we suppress indices most of the time)
\eq{
  \label{matrix_001}
  \arraycolsep2pt
  \begin{array}{rcl@{\hspace{40pt}}rcl}
  \mathcal M &= &\displaystyle \int_{\mathcal X} 
  \left(\arraycolsep2pt \begin{array}{r}\alpha \\-\beta \end{array}\right) \wedge\star
 \left(\arraycolsep2pt \begin{array}{r@{\hspace{2pt}}c@{\hspace{2pt}}l}\alpha &,& -\beta \end{array}\right) ,
  & \mathcal M^T &=& + \mathcal M \,,
  \\[12pt]
  \eta &= &\displaystyle \int_{\mathcal X} 
  \left(\arraycolsep2pt \begin{array}{r}\alpha \\-\beta \end{array}\right) \wedge\hphantom{\star}
 \left(\arraycolsep2pt \begin{array}{r@{\hspace{2pt}}c@{\hspace{2pt}}l}\alpha &,& -\beta \end{array}\right)
  =
  \left( 
  \arraycolsep2pt
  \begin{array}{cc} 0 & -1 \\[2pt] +1 & 0 \end{array}\right),
  & \eta^T &=& - \eta \,,
  \end{array}
}
where $\mathcal M$ is positive definite. 
We also expand the three-forms $F_3$ and $\Xi_A$  introduced in 
\eqref{fluxes_072} and \eqref{hi_001} into $\{\alpha_I,\beta^I\}$ as
\eq{
  \label{matrix_002}
  \arraycolsep2pt
  \begin{array}{lcl@{\hspace{50pt}}lcl}
  F_3 &=&  \displaystyle  \left(\arraycolsep2pt \begin{array}{r@{\hspace{2pt}}c@{\hspace{2pt}}l}\alpha &,& -\beta \end{array}\right) \hat F_3 \,,
  &
  \hat F_3 &=&\displaystyle  \binom{F_{3\op}{}^I}{F_{3\op I}} \,,
  \\[14pt]
  \Xi_A  &=&  \displaystyle   \left(\arraycolsep2pt \begin{array}{r@{\hspace{2pt}}c@{\hspace{2pt}}l}\alpha &,& -\beta \end{array}\right) \hat \Xi_A \,,
  &
  \hat \Xi_A &=& \displaystyle \binom{\Xi^I{}_A}{\Xi_{IA}}\,,
  \end{array}
}
and to avoid confusion let us state that vector $\hat F_3$ has $d_1$ integer components and that 
 $\hat \Xi$ is a $d_1\times d_2$ dimensional matrix with integer components.

%%%%%%%%%%%%%%%%%%%%%%%%%%%%%%%%%%%%%%%%%%%%%%%
%%%%%%%%%%%%%%%%%%%%%%%%%%%%%%%%%%%%%%%%%%%%%%%

\subsubsection*{Stabilizing the $c^A$}

We now turn to the stabilization of the real parts $c^A$ of the K\"ahler-sector moduli 
$\mathsf T^A$. 
The K\"ahler potential $K_{\rm K}$ shown in \eqref{lu_005}, 
its derivatives $K_A=\partial_A K$ as well as the K\"ahler metric $G_{A\ov B}$ do 
not depend on the $c^A$, and in the F-term potential \eqref{hi_012}
we can isolate the $c^A$ contributions as follows
\eq{
 V_{\rm F}& = \frac{e^{K_{\rm K}}}{2}
 \int_{\mathcal X} \bigl(F_3 - \Xi_A c^A\bigr) \wedge \star \bigl(F_3 - \Xi_B c^B \bigr) + \ldots
 \\
& = \frac{e^{K_{\rm K}}}{2}\, \bigl( \hat F^T_3 - c^T \Xi^T\bigr)\, \mathcal M \,
 \bigl( \hat F_3 -  \Xi\op c\bigr) + \ldots\,,
 }
where we used the first Bianchi identity in \eqref{bianchi_001} and 
employed the matrix notation introduced in \eqref{matrix_001} and \eqref{matrix_002}.
By similar  arguments we see that the D-term potential 
\eqref{pot_001} does not depend on $c^A$. 
The extremum of the combined scalar potential for the $c^A$ is then determined as
\eq{
  0 = \partial_{c^A} V  \hspace{40pt}\Longrightarrow\hspace{40pt}
  0=  \Xi^T \hspace{-1pt}\mathcal M \op \hat F_3 - (\op\Xi^T \hspace{-1pt}\mathcal M \,\Xi\op) \op c \,,
}
and in order to stabilize all of the $c^A$ the $d_2\times d_2$ matrix 
$\Xi^T\hspace{-1pt} \mathcal M \,\Xi$ has to have maximal rank equal to $d_2$.
Since $\mathcal M$ is a  $d_1\times d_1$-dimensional matrix, 
we therefore obtain the restriction
\eq{
  \label{hn_001}
  d_2 \leq d_1 \,.
}

%%%%%%%%%%%%%%%%%%%%%%%%%%%%%%%%%%%%%%%%%%%%%%%
%%%%%%%%%%%%%%%%%%%%%%%%%%%%%%%%%%%%%%%%%%%%%%%

\subsubsection*{Bianchi identities I}

We can however make \eqref{hn_001} more precise by taking into account the 
Bianchi identities for the fluxes. 
To do so, we
first perform a singular value decomposition of the flux matrix $\hat \Xi$ as
\eq{
  \label{svd_003}
  \hat \Xi = U\op \Sigma \op V^T \,,
  \hspace{40pt}
  \Sigma = \left( 
  \arraycolsep2pt
  \begin{array}{c} \sigma \\ 0 \end{array} \right),
}
where $U$ is a $d_1 \times d_1$ dimensional orthogonal matrix,
$V$ is a $d_2 \times d_2$ dimensional orthogonal matrix 
and $\sigma$ is a $d_2 \times d_2$-dimensional diagonal matrix 
with the singular values of $\hat \Xi$ on the diagonal. Note that 
since we require $(\op\Xi^T\hspace{-1pt} \mathcal M \,\Xi\op)$ to have maximal rank, 
$\sigma$ has to have maximal rank and therefore  is invertible. 
Turning then to the first Bianchi identity in \eqref{bianchi_001}, we see that using matrix notation we can write
\begin{align}
  \label{svd_001b}
  0 = \hat \Xi^T \eta \: \hat \Xi
   \hspace{40pt}&\Longrightarrow\hspace{40pt}
   0 = V  \left(\arraycolsep2pt \begin{array}{r@{\hspace{2pt}}c@{\hspace{2pt}}l}\sigma &,& 0 \end{array}\right)
   U^T \eta\: U   \arraycolsep2pt
  \left( \begin{array}{c} \sigma \\ 0 \end{array} \right) V^T \,,
  \\[8pt]
  \label{svd_001}
  &\Longrightarrow\hspace{40pt}U^T \eta\: U = \left( \begin{array}{cc} 0 & B \\[2pt] -B^T & C \end{array}\right) ,
\end{align}
where the upper-left block of $U^T \eta\: U$ has dimensions $d_2\times d_2$ and $B$ has dimensions
$d_2\times (d_1-d_2)$. Now, the determinant of the left-hand side in \eqref{svd_001} is equal to one, 
and hence also the right-hand side has to have a non-vanishing determinant (equal to one). 
This is however only possible if $(d_1-d_2)\geq d_2$, that is $d_2\leq d_1/2$, which means
\eq{
  \label{hn_002}
  \bom{
  h^{1,1}_{\vphantom{-}}\leq h^{2,1}_- \,.
  }
}
Hence, in order to stabilize all closed-string moduli by fluxes and satisfy the 
Bianchi identities, a necessary requirement is given by \eqref{hn_002}.

%%%%%%%%%%%%%%%%%%%%%%%%%%%%%%%%%%%%%%%%%%%%%%%
%%%%%%%%%%%%%%%%%%%%%%%%%%%%%%%%%%%%%%%%%%%%%%%

\subsubsection*{Bianchi identities II}

Let us also consider the third Bianchi identity in \eqref{bianchi_001}. Expanding the 
three-form $\Theta^A$ in the basis $\{\alpha_{\lambda},\beta^{\lambda}\}\in H^3_+(\mathcal X)$
similarly as in \eqref{matrix_002} gives rise to a matrix $\hat \Theta$ of dimensions
$2\op h^{2,1}_+ \times (h^{1,1}+1)$. Using matrix notation we then have
\eq{
  0 = \hat \Xi \, \hat \Theta^T
  \hspace{30pt}\Longrightarrow\hspace{30pt}
  0 = U\op \Sigma \op V^T \op \hat\Theta^T
  \hspace{30pt}\Longrightarrow\hspace{30pt}
  0 =\hat\Theta^T \,,
}
where we used that $\hat\Xi$ has maximal rank and hence the diagonal matrix $\sigma$ in 
$\Sigma$ is invertible. We can therefore conclude that 
when stabilizing all closed-string moduli by fluxes, Bianchi identities imply that $\Theta^A=0$ and 
therefore the D-term potential vanishes. 
Without loss of generality, for the purpose of moduli stabilization we can therefore restrict ourselves to 
\eq{
  \bom{
  h^{2,1}_+ = 0 \,.
  }
}

%%%%%%%%%%%%%%%%%%%%%%%%%%%%%%%%%%%%%%%%%%%%%%%
%%%%%%%%%%%%%%%%%%%%%%%%%%%%%%%%%%%%%%%%%%%%%%%
%%%%%%%%%%%%%%%%%%%%%%%%%%%%%%%%%%%%%%%%%%%%%%%
%%%%%%%%%%%%%%%%%%%%%%%%%%%%%%%%%%%%%%%%%%%%%%%
%%%%%%%%%%%%%%%%%%%%%%%%%%%%%%%%%%%%%%%%%%%%%%%
%%%%%%%%%%%%%%%%%%%%%%%%%%%%%%%%%%%%%%%%%%%%%%%

\section{The minimal case: $h^{2,1}_-=h^{1,1}$}
\label{sec_minimal_case}

We now consider the situation $h^{2,1}_-=h^{1,1}_{\vphantom{+}}$ and $h^{2,1}_+=0$, 
which is the minimal 
case for stabilizing all closed-string moduli by fluxes. In this section we discuss the following questions:
\begin{itemize}

\item In section~\ref{sec_manifest_tadpole} we rewrite  the scalar potential and replace the  
R-R three-form flux $F_3$ by the tadpole charges $N_A$. This formulation is suitable 
for performing computer-based scans for flux vacua. 

\item In section~\ref{sec_no_susy_mink} we show that supersymmetric Minkowski vacua 
with all moduli stabilized by fluxes are always singular and should be excluded,
and

\item in section~\ref{sec_stab_rad} we discuss moduli stabilization of the 
radial K\"ahler-sector modulus and determine restrictions for de-Sitter vacua. 

\end{itemize}

%%%%%%%%%%%%%%%%%%%%%%%%%%%%%%%%%%%%%%%%%%%%%%%
%%%%%%%%%%%%%%%%%%%%%%%%%%%%%%%%%%%%%%%%%%%%%%%
%%%%%%%%%%%%%%%%%%%%%%%%%%%%%%%%%%%%%%%%%%%%%%%
%%%%%%%%%%%%%%%%%%%%%%%%%%%%%%%%%%%%%%%%%%%%%%%

\subsection{Manifest tadpole charges}
\label{sec_manifest_tadpole}

We start by showing how, after the moduli $c^A=\mbox{Re}\,\mathsf T^A$ have been stabilized, 
the R-R three-form flux $F_3$ can be replaced by the tadpole charges $N_A$. 
Such a rewriting is useful for eliminating the dependence on $F_3$ in 
computer-based searches for flux vacua.

%%%%%%%%%%%%%%%%%%%%%%%%%%%%%%%%%%%%%%%%%%%%%%%
%%%%%%%%%%%%%%%%%%%%%%%%%%%%%%%%%%%%%%%%%%%%%%%

\subsubsection*{Singular value decomposition}

Let us recall and expand our discussion from  section~\ref{sec_hodge_numbers}.
In equation \eqref{svd_003}  we have performed a singular value decomposition of the flux matrix 
$\hat \Xi$  as
\eq{
  \label{svd_003b}
  \hat \Xi = U\op \Sigma \op V^T \,,
  \hspace{40pt}
  \Sigma = \left( 
  \arraycolsep2pt
  \begin{array}{c} \sigma \\ 0 \end{array} \right),
}
where $\sigma$ is an invertible diagonal matrix. 
Here we specialize to the 
situation $h^{1,1}=h^{2,1}_-$, and for ease of presentation we use $d=h^{1,1}+1=h^{2,1}_-+1$.
The orthogonal $2d\times 2d$ matrix $U$ can be expressed in terms of two 
$2d\times d$ blocks $u_{1}$ and $u_2$, and we introduce the notation 
\eq{
  U = (u_1,u_2) \,, \hspace{30pt} 
  U^T \hat F_3 = \arraycolsep1pt\left( \begin{array}{c}f^1\\f^2\end{array}\right),  \hspace{30pt}
  U^T \mathcal M \op U = 
  \arraycolsep2pt
  \left(
  \begin{array}{cc} \mathsf M_{11} & \mathsf M_{12} \\ \mathsf  M_{21} &  \mathsf M_{22} \end{array}\right),
}
where the $2d\times 2d$ matrix $\mathcal M$ was defined in \eqref{matrix_001}
and the $2d$ vector $\hat F_3$ was defined in \eqref{matrix_002}.
Note  that since $\mathcal M$ is positive definite, also $ \mathsf M_{11}$ is positive definite. 
Furthermore, we have  argued before that $\hat \Xi$ has to satisfy the Bianchi identity
$0 = \hat \Xi^T \eta \: \hat \Xi$, which puts restrictions on $u_{1}$ and $u_2$. 
In particular, in the present situation we find that 
\eq{
\label{svd_bianchi_001}
U^T \eta\, U = \left( \begin{array}{cc} 0 & b \\[2pt] -b^T & 0 \end{array}\right) ,
\hspace{60pt}
\det b = \pm 1\,,
}
where $b= u_1^T\op\eta \op u_2$ is an invertible $d\times d$ matrix,
and the lower-right block vanishes by requiring $(U^T \eta\, U)^2= \mathds 1$.
With a bit of algebra we can then compute
\eq{
  \label{rw_838}
  &\int_{\mathcal X}  \mbox{Re}\, G \wedge\star  \mbox{Re}\,   G 
  = \mathsf F^T  \mathsf M_{11} \op \mathsf F + N^T m^{-1} N \,,
  \\
  &\int_{\mathcal X}  \mbox{Im}\, G \wedge\star  \mbox{Im}\,   G = \tau^T m \,\tau \,,
}
where $N_A$ denotes again the vector of tadpole charges, where $\tau^A = \mbox{Im}\,\mathsf T^A$ are 
the imaginary parts of the combined K\"ahler-sector  moduli, and where we defined
the $d\times d$ matrix and $d$-vector
\eq{
  \label{min_300}
  m = \Xi^T \hspace{-1pt}\mathcal M \,\Xi\,, 
  \hspace{60pt}
  \mathsf F = f^1 +  \mathsf M_{11}^{-1}  \mathsf M_{12}^{\vphantom{-1}} f^2 - \sigma\, V^T  c \,.
}

%%%%%%%%%%%%%%%%%%%%%%%%%%%%%%%%%%%%%%%%%%%%%%%
%%%%%%%%%%%%%%%%%%%%%%%%%%%%%%%%%%%%%%%%%%%%%%%

\subsubsection*{Stabilizing the $c^A$}

Next, we  turn to the scalar F-term potential shown in \eqref{hi_012}. 
As mentioned already in section~\ref{sec_hodge_numbers},
the moduli $c^A=\mbox{Re}\, \mathsf T^A$ only appear through $\mbox{Re}\, G$, 
and using \eqref{rw_838} we obtain
\eq{
\label{rw_839a}
V_F = \frac{e^{K_{\rm K}}}{2} \, \int_{\mathcal X} 
   \mbox{Re}\, G \wedge\star  \mbox{Re}\,   G  + \ldots
   = \frac{e^{K_{\rm K}}}{2} \, \mathsf F^T  \mathsf M_{11} \op \mathsf F + \ldots\,.
}
Since by assumption the $d\times d$ matrix $\mathsf M_{11}$ has maximal rank $d$, minimizing
\eqref{rw_839a} with respect to $c^A$ fixes all  $c^A$ in terms of the flux matrix $\hat\Xi$ and complex-structure moduli as
\eq{
  \label{rw_839}
  \mathsf F = 0 
  \hspace{40pt}\Longrightarrow\hspace{40pt}
  c= V \sigma^{-1} \bigl(  f^1 + \mathsf M_{11}^{-1} \mathsf M^{\vphantom{-1}}_{12} f^2 \bigr) \,. 
}
However,  for our purpose the precise 
values of the stabilized $c^A$ are not important. 
We furthermore note that with $\phi$ denoting all real moduli except the $c^A$, the mass matrix takes the form
\eq{
 M^2 = 
 \arraycolsep2pt
 \left[ 
 \begin{array}{cc}
 e^{K_{\rm K}}m &   0 \\
 0 & M^2_{\phi\phi}
 \end{array}
 \right],\hspace{60pt}
 m =  \Xi^T \hspace{-1pt}\mathcal M \,\Xi\,,
}
where $m$ is positive definite and 
$M^2_{\phi\phi}$ denotes the mass matrix for all moduli $\phi$. Hence, the stabilization 
of the $c^A$ can be separated from the problem of stabilizing the remaining moduli.

%%%%%%%%%%%%%%%%%%%%%%%%%%%%%%%%%%%%%%%%%%%%%%%
%%%%%%%%%%%%%%%%%%%%%%%%%%%%%%%%%%%%%%%%%%%%%%%

\subsubsection*{Scalar potential}

We finally turn to the full form of the F-term potential \eqref{hi_012}
after the $c^A$ moduli have been stabilized. We denote this potential by 
$\tilde V_F=V_F\op\rvert_{{\rm min}\,c^A}$. 
Separating the three-form flux $G$ into real and imaginary components,
using the first Bianchi identity in \eqref{bianchi_001} and 
employing equation \eqref{rw_838}, this potential can be written as
\eq{
   \label{pot_943}
  \bom{
  \tilde V_F 
   = e^{K_{\rm K}} \biggl[\,
\frac{1}{2} \op N^T m^{-1} N + N^T \tau + \frac{1}{2} \op\tau^T m \,\tau
+  e^{-K_{\rm cs}}\, \Xi^{\ov 0}_{A\vphantom{\ov B}} \, \omega^{A\ov B}\,\Xi^0_{\ov B}\,
   \biggr] \,.
   }
}
From here we see that indeed, after the $c^A$ have been stabilized, we can express the F-term potential 
using the tadpole charges $N_A$ instead of the R-R three-form flux $F_3$.
However, of course one has to check whether for a given $N$ and flux matrix $\hat \Xi$ 
there exists a corresponding $\hat F_3$ with integer components such that $N = \hat\Xi^T \eta \op \hat F_3$.
Coming then back to the expression for the scalar potential at the minimum derived in 
\eqref{hi_011},  using \eqref{rw_838} together with \eqref{rw_839} we
find
\eq{
   V\op\bigr\rvert_{\rm extr} = -\frac{e^{K_{\rm K}}}{2} \Bigl[  \op
   N^T m^{-1}  N + 
  N^T \tau \Bigr]_{\rm extr}\,,
}
where we recall that the matrix $m = \Xi^T \hspace{-1pt}\mathcal M \,\Xi$ was defined in \eqref{min_300}.
We discuss this expression in more detail below.

%%%%%%%%%%%%%%%%%%%%%%%%%%%%%%%%%%%%%%%%%%%%%%%
%%%%%%%%%%%%%%%%%%%%%%%%%%%%%%%%%%%%%%%%%%%%%%%
%%%%%%%%%%%%%%%%%%%%%%%%%%%%%%%%%%%%%%%%%%%%%%%
%%%%%%%%%%%%%%%%%%%%%%%%%%%%%%%%%%%%%%%%%%%%%%%

\subsection{No supersymmetric Minkowski vacua}
\label{sec_no_susy_mink}

We now want to show that in the case $h^{2,1}_-=h^{1,1}_{\vphantom{+}}$,
supersymmetric Minkowski vacua  require 
the complex-structure moduli to be stabilized at the boundary 
of  moduli space. Such solutions should therefore be excluded. 
Note that for type IIA compactifications with non-geometric fluxes a similar result 
has been derived in \cite{Micu:2007rd}.

%%%%%%%%%%%%%%%%%%%%%%%%%%%%%%%%%%%%%%%%%%%%%%%
%%%%%%%%%%%%%%%%%%%%%%%%%%%%%%%%%%%%%%%%%%%%%%%

\subsubsection*{Analysis}

To show the above statement, we recall the general form of the F-term potential \eqref{pot_003} and note 
that for supersymmetric Minkowski vacua we have to require
\eq{
  \label{susy_001}
  0 = W \,, \hspace{40pt} 0 = F_A = \partial_A W
  \,, \hspace{40pt} 0 = F_i = \partial_i W \,,
}
where $A=0,\ldots, h^{1,1}$ labels the combined K\"ahler-sector moduli and $i=1,\ldots, h^{2,1}_-$ 
labels the complex-structure moduli. The superpotential $W$ was defined
in \eqref{pot_002}, for which the second condition in \eqref{susy_001}
reads
\eq{
  \label{susy_002}
  0 = \int_{\mathcal X} \Xi_A \wedge \Omega\,.
}
Let us now expand the holomorphic three-form $\Omega$ 
into the symplectic basis 
\eqref{basis_001} and define
\eq{
  \Omega = (\alpha,-\beta)\, \binom{\mathsf X}{\mathsf F} \,,
  \hspace{40pt}
  U^T \binom{\mathsf X}{\mathsf F} = \binom{x^1}{x^2} \,,
  }
where $\mathsf X$ and $\mathsf F$ denote the periods of the Calabi-Yau three-fold, 
the $2d\times 2d$ matrix $U$ appeared in the singular value decomposition 
\eqref{svd_003b} and where we suppressed indices. 
Using this notation together with the expansion of $\Xi_A$ given in \eqref{matrix_002}, 
the singular value decomposition of $\hat \Xi$ shown in \eqref{svd_003b} and
the form of $U^T\eta \, U$ given in \eqref{svd_bianchi_001} allows us to write \eqref{susy_002} as
\eq{
  0 = \int_{\mathcal X} \Xi\wedge \Omega = V \op \bigl(\sigma,0\bigr) \,U^T \eta\, U \,U^T \binom{\mathsf X}
  {\mathsf F} = 
  V \op\sigma\op b \op x^2
  \hspace{20pt}\Longrightarrow \hspace{20pt} x^2 = 0\,,
}
where we used that the $d\times d$ matrices $V$, $\sigma$ and $b$ are invertible. 
We can now use this solution to evaluate
\eq{
  \int_{\mathcal X} \Omega\wedge \ov\Omega = \bigl( \mathsf X,\mathsf F \bigr) \,\eta\, \binom{\ov{\mathsf{X}}}{\ov{\mathsf {F}}} = 
  \bigl( x^{1T}, x^{2T} \bigr) \left( \begin{array}{cc} 0 & b \\[2pt] -b^T & 0 \end{array}\right) 
  \binom{\ov x{}^1}{\ov x{}^2}
  \overset{x^2=0}{=} 0\,.
}
This implies that the K\"ahler potential $K_{\rm cs}$ shown in \eqref{lu_005}
is singular, and hence the complex-structure moduli are stabilized at the boundary of their moduli space.
Note that since we did not specify the form of the periods this result holds for the
quantum-corrected periods, and we have to exclude such situations.

%%%%%%%%%%%%%%%%%%%%%%%%%%%%%%%%%%%%%%%%%%%%%%%
%%%%%%%%%%%%%%%%%%%%%%%%%%%%%%%%%%%%%%%%%%%%%%%

\subsubsection*{Remark}

We close this subsection with the following remark.
For $h^{1,1}_{\vphantom{-}}<h^{2,1}_-$ the argument presented above is modified 
and supersymmetric Minkowski vacua stabilizing all moduli cannot be excluded.
The condition \eqref{susy_002} is equal to $\Xi_A^{\ov 0}=0$, which 
when inserted in \eqref{hi_012b} gives the condition 
\eq{
  \mbox{supersymmetric Minkowski vacuum:}\hspace{40pt} \star G =  i \op G \,.
}
This is the well-known requirement of $G$ being imaginary self-dual for the case of 
only $H$-flux present
\cite{Giddings:2001yu}.

%%%%%%%%%%%%%%%%%%%%%%%%%%%%%%%%%%%%%%%%%%%%%%%
%%%%%%%%%%%%%%%%%%%%%%%%%%%%%%%%%%%%%%%%%%%%%%%
%%%%%%%%%%%%%%%%%%%%%%%%%%%%%%%%%%%%%%%%%%%%%%%
%%%%%%%%%%%%%%%%%%%%%%%%%%%%%%%%%%%%%%%%%%%%%%%

\subsection{Stabilizing the radial K\"ahler-sector modulus}
\label{sec_stab_rad}

In this section we study moduli stabilization and stability conditions for 
the radial K\"ahler-sector modulus, and discuss conditions for obtaining de-Sitter 
minima.

%%%%%%%%%%%%%%%%%%%%%%%%%%%%%%%%%%%%%%%%%%%%%%%
%%%%%%%%%%%%%%%%%%%%%%%%%%%%%%%%%%%%%%%%%%%%%%%

\subsubsection*{Moduli stabilization}

To do so, we first split the combined K\"ahler-sector  modulus $\tau^A$ 
and the tadpole-charge vector $N_A$ into radial and 
angular parts, that is, using the Euclidean norm $\delta\equiv \delta_{AB}$ and suppressing 
indices we define
\eq{
  \arraycolsep2pt
  \begin{array}[c]{lcl}
  \tau  &=& \mathsf r \, e_{\tau} \,, 
  \\[4pt]
  N &=& \mathsf n\, e_{N} \,,
  \end{array}  
  \hspace{40pt}\mbox{where}\hspace{40pt} 
  \arraycolsep2pt
  \begin{array}[c]{r@{\op}l@{\op}lcl@{\hspace{25pt}}r@{\op}l@{\op}lcl}
  \tau^T &\delta &\tau &=& \mathsf r^2 \,, & \ e_{\tau}^T &\delta & e^{\vphantom{T}}_{\tau} &=& 1\,,
  \\[4pt]
  N^T& \delta^{-1}& N &=& \mathsf n^2 \,, & \ e_{N}^T &\delta^{-1} & e^{\vphantom{T}}_{N} &=& 1\,.
  \end{array}
}
Our convention is  $\mathsf r\geq0$ and $\mathsf n\geq0$. 
Next, we note that the K\"ahler potential $K_{\rm K}$
and the matrix $\omega^{A\ov B}$ 
appearing in the scalar potential \eqref{pot_943}  depend on the radial modulus 
$\mathsf r$ as $e^{K_{\rm K}}\sim \mathsf r^{-4}$ 
and $\omega \sim \mathsf r^2$. We can then express the scalar potential
\eqref{pot_943}  as
\eq{
  \label{pot_090}
  \tilde V_F = \frac{1}{\mathsf r^4} \left[\, \frac{a}{2} + b\, \mathsf r + \frac{c}{2} \op \mathsf r^2 \,\right] \,,
}
where the coefficients $a,b,c$ are independent of $\mathsf r$. 
Using $e^{\tilde K_{\rm K}} = \mathsf r^4 \op e^{K_{\rm K}}$ and $\tilde \omega^{A\ov B} = \mathsf r^{-2} \omega^{A\ov B}$, 
they are given by 
\eq{
 \arraycolsep2pt
 \begin{array}[c]{lcl}
 a &=& \displaystyle e^{\tilde K_{\rm K}} \op \mathsf n^2 \, e_N^T m^{-1} e_N \,, 
 \\[4pt]
 b &=& \displaystyle e^{\tilde K_{\rm K}}\op \mathsf n^{\hphantom{2}} \, e_N^T \, e_{\tau}\,, 
 \end{array}
 \hspace{60pt}
 c =e^{\tilde K_{\rm K}} \Bigl(  e_{\tau}^T m \, e_{\tau}^{\vphantom{T}}
 + 2\op  e^{-K_{\rm cs}}\, \Xi^{\ov 0}\, \tilde \omega\,\Xi^0 \Bigr)\,,
}
where the matrix $m$ was defined in \eqref{min_300},
and since  $m$ is positive definite we have $a\geq 0$.
We can then minimize the potential \eqref{pot_090} with respect to $\mathsf r$ by solving 
$\partial_{\mathsf r} \tilde V_F=0$. Imposing in addition the necessary condition 
$\partial_{\mathsf r}^2 \tilde V_F >0$
for the minimum to be stable, we obtain the following necessary  conditions for stable minima:
\eq{
  \label{minima_832}
  \renewcommand{\arraystretch}{1.45}
  \arraycolsep10pt
  \begin{array}{|l@{\hspace{50pt}}lll|}
  \hline
  \multirow{2}{*}{\mbox{AdS minimum}} & a>0 & b\leq 0 & c<\frac{b^2}{a} 
  \\
  & a>0 & b> 0 & c<0 
  \\
  \hline
  \mbox{Minkowski minimum} & a>0 & b<0 & c=\frac{b^2}{a} 
  \\
  \hline
  \mbox{dS minimum} & a>0 & b<0 & \frac{b^2}{a} < c < \frac{9}{8} \op\frac{b^2}{a}
  \\
  \hline
  \end{array}
}  
Note however that our requirement \eqref{calib_007} excludes  AdS minima with $b>0$. 
Furthermore,  when ignoring $\alpha'$-corrections to the K\"ahler potential and only 
considering $H$-flux, the term $\Xi^{\ov 0}\, \tilde \omega\,\Xi^0$ vanishes and 
$c>0$.

%%%%%%%%%%%%%%%%%%%%%%%%%%%%%%%%%%%%%%%%%%%%%%%
%%%%%%%%%%%%%%%%%%%%%%%%%%%%%%%%%%%%%%%%%%%%%%%

\subsubsection*{Stable de-Sitter minima}

Let us now focus on stable de-Sitter minima. As we can see from \eqref{minima_832},
after $\mathsf r$ has been stabilized the remaining moduli have to be fixed
in a rather restricted region. 
This is in agreement with the observation made in \cite{Danielsson:2012by,Blaback:2013ht} that 
stable de-Sitter vacua are located in a narrow band in moduli space. 
For the stabilized radial modulus $\mathsf r$ the constraints on $c$ shown in \eqref{minima_832}
translate into 
\eq{
  \mbox{dS minimum:} \hspace{40pt}
  \frac{a}{|b|}< \mathsf r_{\rm min} <  \frac{4}{3}\, \frac{a}{|b|} 
  \hspace{25pt}\mbox{with}\hspace{25pt}
  \frac{a}{|b|} = \mathsf n \, \frac{e_N^T \op m^{-1} e_N}{| e_N^T \, e_{\tau}|} \,,
}
and  from here we see that a large value for $\mathsf r_{\rm min}$ -- necessary 
for a large-volume and small-coupling limit -- can be achieved 
in three different ways:
\begin{enumerate}

\item The first possibility to make $a/|b|$ large is to require 
$\mathsf n\gg1$. This means that the tadpole charges have to be large, 
requiring a large number of O3- and O7-planes to be present. 
Such configurations exists -- but these usually have  
a complicated topology and hence a large number of moduli have to be stabilized.
Problems related to the latter issue have been discussed recently for instance in 
\cite{Gao:2020xqh,Braun:2020jrx,Bena:2020xrh}.

\item A second possibility for obtaining a large $\mathsf r_{\rm min}$ is to 
stabilize the complex-structure moduli such that $e_N^T\op m^{-1} e_N\gg 1$,
where $m$ was defined in \eqref{min_300}. Since $e_N$ is normalized to one, such 
a condition  is possible only near a boundary of complex-structure moduli space.

\item A third way to make $a/|b|$ large is to have 
$| e_N^T \, e_{\tau}|\ll 1$. Demanding the moduli to be stabilized in a large-volume and weak-coupling regime 
implies that all components of $\tau^A$  have to be large (excluding the $h^{1,1}_-$ moduli). This means
that approximately we have \raisebox{0pt}[0pt][-20pt]{$\displaystyle e_{\tau} \sim (1,1,\ldots, 1)/\sqrt{h^{1,1}}$}.
Since we require all components of the tadpole charges to be non-positive, to achieve 
$| e_N^T \, e_{\tau}|\ll 1$ we therefore may look for compactifications with $e_N = (0,\ldots,0,-1,0,\ldots)$ and 
$h^{1,1}\gg 1$. 

\end{enumerate}
To summarize, for stable de-Sitter vacua the moduli have to be stabilized in 
a narrow region of moduli space and one typically has to consider  non-generic 
situations. 
This explains why it is difficult to find de-Sitter solutions, however, from our analysis here we do not 
see any reason why they cannot exist.

%%%%%%%%%%%%%%%%%%%%%%%%%%%%%%%%%%%%%%%%%%%%%%%
%%%%%%%%%%%%%%%%%%%%%%%%%%%%%%%%%%%%%%%%%%%%%%%
%%%%%%%%%%%%%%%%%%%%%%%%%%%%%%%%%%%%%%%%%%%%%%%
%%%%%%%%%%%%%%%%%%%%%%%%%%%%%%%%%%%%%%%%%%%%%%%
%%%%%%%%%%%%%%%%%%%%%%%%%%%%%%%%%%%%%%%%%%%%%%%
%%%%%%%%%%%%%%%%%%%%%%%%%%%%%%%%%%%%%%%%%%%%%%%

\section{Comments on de-Sitter vacua}
\label{sec_desitter}

A number of approaches for constructing de-Sitter vacua using non-geometric fluxes have
appeared in the literature
\cite{
deCarlos:2009fq, 
deCarlos:2009qm, 
Dibitetto:2010rg, 
Dibitetto:2011gm,
Danielsson:2012by,
Blaback:2013ht, 
Damian:2013dq,
Damian:2013dwa,
Blaback:2015zra,
Shukla:2016xdy,
Cribiori:2019hrb,
CaboBizet:2020cse
}.
We briefly review them in this section.
For the stable constructions which in addition to the R-R three-form flux $F_3$ only use 
(non-)geometric
\mbox{$H$-}, $F$-, $Q$- and $R$-fluxes
to stabilize all closed-string moduli,
we argue that none of them satisfy all the consistency conditions discussed above.

%%%%%%%%%%%%%%%%%%%%%%%%%%%%%%%%%%%%%%%%%%%%%%%
%%%%%%%%%%%%%%%%%%%%%%%%%%%%%%%%%%%%%%%%%%%%%%%

\subsubsection*{The de Carlos-Guarino-Moreno model}

The authors of \cite{deCarlos:2009fq,deCarlos:2009qm} 
have constructed a family of stable de-Sitter minima for type IIB string theory 
compactified on the isotropic 
$\mathbb T^6/\mathbb Z_2\times \mathbb Z_2$ 
orientifold with O3-/O7-planes in the presence of geometric and non-geometric fluxes. 
Let us recall a particular representative of this family using our conventions:
\begin{itemize}

\item The isotropic torus can be seen as a compactification manifold 
with topology characterized by $h^{1,1}_+=h^{2,1}_-=1$ and $h^{1,1}_-=h^{2,1}_+=0$. 
Up to an irrelevant additive constant, the corresponding K\"ahler potential in the 
large-volume, small-coupling and large-complex-structure limit is  given by
\eq{
  \label{cgm_001}
  K=  -\log\bigl[ - i\op (\mathsf T^0-\ov{\mathsf{T}}{}^0\bigr)\bigr]  
   -3\log\bigl[ - i\op (\mathsf T^1-\ov{\mathsf{T}}{}^1\bigr)\bigr]
   -3\log\bigl[ - i\op (z^1-\ov{z}{}^1\bigr)\bigr].
}

\item The R-R three-form flux, the NS-NS $H$- and $Q$-flux and the tadpole charges 
can be expressed using the matrix notation introduced in \eqref{matrix_002}. In particular,
we have
\eq{
  \label{cgm_003}
  \hat F_3 = 
  \left( \arraycolsep1pt \begin{array}{r}
  -47
  \\
  -44
  \\
  44
  \\
  51
  \end{array}
  \right),
  \hspace{40pt}
  \hat \Xi = 
  \left( \arraycolsep1pt \begin{array}{r@{\hspace{8pt}}r}
  44 & 0 \\ 
  1 & -1 \\
  -1 & -3 \\
  132 & 0
  \end{array}
  \right),
  \hspace{40pt}
  N = 
    \left( \arraycolsep1pt \begin{array}{r}
  -7748
  \\
  192
  \end{array}
  \right),
}
where the first column in $\hat \Xi$ corresponds to the $H$-flux and 
the second column to the non-geometric $Q$-flux. 
Note also that the  fluxes contained in $\hat \Xi$ satisfy the Bianchi identity \eqref{bianchi_001}
and that $\hat \Xi$ is of maximal rank equal to $2$.

\item Using this data for instance in  \eqref{pot_943} and minimizing 
the resulting potential, the authors of 
\cite{deCarlos:2009fq,deCarlos:2009qm} find a  de-Sitter minimum with $V\rvert_{\rm min}=1.43\cdot 10^{-3}$  at
\eq{
  \label{cgm_002}
  \tau^0_{\rm min} = 2.008\,, \hspace{30pt}
  \tau^1_{\rm min} = 48.684\,, \hspace{30pt}
  z^1_{\rm min} = -1.034+1.144i \,.
}
(As explained above, the $c^A$ dependence can be decoupled.)
This minimum is stable, and the eigenvalues of the corresponding ca\-non\-i\-cally-normalized
mass matrix are
\eq{
  \mbox{eigenvalues of $M^2_{\rm can}$\,:}\hspace{20pt}
  \bigl\{ \,4.25\op, \; 1.75\op, \; 1.46 \op, \; 2.29\cdot 10^{-3}\,\bigr\}\,.
}

\end{itemize}
This example illustrates  that the supergravity equations following from 
non-geo\-me\-tric flux compactifications can 
have stable de-Sitter solutions. 
However, we believe that
this model is not consistent in string theory: 
in section~\ref{sec_tadpole_charges} we argued that 
in order to avoid arbitrary-large gauge groups
the tadpole charges $N_A$ should be non-positive, and 
from \eqref{cgm_003} we see that the present 
model violates this requirement. (This observation extends to the whole family of models.) We therefore do not 
consider it to be a consistent string-theory construction.

%%%%%%%%%%%%%%%%%%%%%%%%%%%%%%%%%%%%%%%%%%%%%%%
%%%%%%%%%%%%%%%%%%%%%%%%%%%%%%%%%%%%%%%%%%%%%%%

\subsubsection*{Other de-Sitter constructions with non-geometric fluxes I}

Other papers known to us which construct de-Sitter vacua  -- using in addition to the R-R three-form flux $F_3$ 
only (non-)geometric $H$-, $F$-, $Q$- and $R$-fluxes --
 are the following:
\begin{itemize}

\item From a supergravity point of view the gauge algebras 
originating from  non-geometric fluxes haven been 
studied in \cite{Dibitetto:2010rg,Dibitetto:2011gm}. However, no
stable de-Sitter models have been found.

\item In \cite{Danielsson:2012by} a search for  de-Sitter vacua
has been performed for the $\mathbb T^6/\mathbb Z_2\times \mathbb Z_2$
orientifold, but no fully stable de-Sitter extrema without any tachyonic directions 
have been found.

\item In \cite{Blaback:2013ht} the authors found  three stable de-Sitter minima
with the help on an evolutionary algorithm. However, one can check 
that these models (summarized in appendix A of \cite{Blaback:2013ht}) 
do not satisfy the Bianchi identities \eqref{bianchi_001}.
We therefore do not consider them to be consistent string-theory models.

\item In \cite{Damian:2013dq} as well as in \cite{Damian:2013dwa}
the authors considered compactifications 
of type IIB string theory on  the 
$\mathbb T^6/\mathbb Z_2\times \mathbb Z_2$ 
orientifold with O3-/O7-planes and  \mbox{(non-)}geometric fluxes. With the help of a genetic algorithm 
a number of stable de-Sitter minima are obtained for this setting,
however, we were not able to reproduce these results.\footnote{We thank C.~Damian and 
O.~Loaiza-Brito for correspondence on this question.}

\item In \cite{Shukla:2016xdy} Bianchi identities for toroidal type IIB compactifications have been analyzed,
but no stable de-Sitter solutions were found.

\item In \cite{CaboBizet:2020cse} the authors consider non-geometric flux compactifications
of type IIB string theory on the isotropic $\mathbb T^6/\mathbb Z_2\times \mathbb Z_2$ 
orientifold. Using machine-learning techniques the authors obtain de-Sitter extrema  which,  however,
have
at least one tachyonic direction. 

\end{itemize}
We conclude that none of the non-geometric de-Sitter models reviewed here are fully-stable and consistent
in string theory. This observation is in agreement with the swampland de-Sitter conjecture 
made in  \cite{Obied:2018sgi}.

%%%%%%%%%%%%%%%%%%%%%%%%%%%%%%%%%%%%%%%%%%%%%%%
%%%%%%%%%%%%%%%%%%%%%%%%%%%%%%%%%%%%%%%%%%%%%%%

\subsubsection*{Other de-Sitter constructions with non-geometric fluxes II}

In addition to the $H$-, $F$-, $Q$- and $R$-fluxes discussed in this paper, one can 
consider
the non-geometric $P$-flux. This flux is the 
the S-dual completion of the $Q$-flux \cite{Aldazabal:2006up}, which 
generates quadratic couplings among the 
K\"ahler-sector moduli in the superpotential. 
\begin{itemize}

\item In \cite{Damian:2013dwa,Blaback:2015zra,Cribiori:2019hrb}
the authors include non-geometric $P$-flux for the construction of de-Sitter vacua. We have not 
considered such fluxes in our work and therefore do not comment on these examples.

\end{itemize}

%%%%%%%%%%%%%%%%%%%%%%%%%%%%%%%%%%%%%%%%%%%%%%%
%%%%%%%%%%%%%%%%%%%%%%%%%%%%%%%%%%%%%%%%%%%%%%%
%%%%%%%%%%%%%%%%%%%%%%%%%%%%%%%%%%%%%%%%%%%%%%%
%%%%%%%%%%%%%%%%%%%%%%%%%%%%%%%%%%%%%%%%%%%%%%%
%%%%%%%%%%%%%%%%%%%%%%%%%%%%%%%%%%%%%%%%%%%%%%%
%%%%%%%%%%%%%%%%%%%%%%%%%%%%%%%%%%%%%%%%%%%%%%%

\section{Summary, conclusions, outlook}
\label{sec_sum}

In this paper we have studied compactifications of type IIB string theory 
on Calabi-Yau orientifolds with O3-/O7-planes and (non-)geometric 
$H$-, $F$-, $Q$- and $R$-fluxes. 
For this setting it is possible to stabilize all closed-string moduli classically without
the need for non-perturbative contributions, and  examples
of stable de-Sitter constructions can be found in the literature.

%%%%%%%%%%%%%%%%%%%%%%%%%%%%%%%%%%%%%%%%%%%%%%%
%%%%%%%%%%%%%%%%%%%%%%%%%%%%%%%%%%%%%%%%%%%%%%%

\subsubsection*{Summary}

Our motivation for this work was to investigate whether non-geometric flux compactifications 
allow for stable de-Sitter vacua.
Although we were neither able to show that de-Sitter vacua cannot be obtained from 
non-geometric fluxes nor able to construct a consistent de-Sitter solution, 
we made progress in understanding non-geometric flux compactifications. More concretely,
\begin{itemize}

\item based on the requirement that the total rank of the gauge group should be bounded from above
(cf.~\cite{Vafa:2005ui}), in section~\ref{sec_tadpole_charges}
we argued that the contributions of closed-string fluxes to the 
tadpole cancellation conditions should be similar to D-branes and not anti-D-branes. 
In our conventions this means $N_A\leq 0$.

\item In section~\ref{sec_mod_stab} we  derived a simple expression for the scalar potential 
at the minimum (cf.~equation \eqref{hi_011}). We furthermore showed that in order to stabilize all closed-string moduli 
classically by $H$-, $F$-, $Q$-, $R$- and $F_3$-fluxes, the topology of the compactification manifold 
has to be restricted as $h^{1,1}_{\vphantom{-}} \leq h^{2,1}_-$ and we can choose $h^{2,1}_+ = 0$
without loss of generality.

\item When specializing to the case $h^{1,1}_{\vphantom{-}} = h^{2,1}_-$, we were able to 
eliminate the dependence on the R-R three-form flux $F_3$ in favor 
of the tadpole charges $N_A$ in the scalar potential (cf.~equation \eqref{pot_943}).
This form promises to be very useful for performing computer-based searches for 
flux vacua. 
We also showed that for $h^{1,1} = h^{2,1}_-$ it is not possible to obtain
supersymmetric Minkowski vacua, and 
we derived constraints for stable de-Sitter vacua. In particular, for the latter we concluded
that moduli have to be stabilized in a rather restricted region of moduli space.

\item In section~\ref{sec_desitter} we reviewed known de-Sitter constructions 
based on non-geo\-me\-tric flux compactifications. 
For the fully-stable constructions with only  $H$-, $F$-, $Q$- and $R$-fluxes 
we concluded that these are not consistent in string theory. 

\end{itemize}

%%%%%%%%%%%%%%%%%%%%%%%%%%%%%%%%%%%%%%%%%%%%%%%
%%%%%%%%%%%%%%%%%%%%%%%%%%%%%%%%%%%%%%%%%%%%%%%

\subsubsection*{Conclusions}

The swampland conjecture made in \cite{Obied:2018sgi} 
proposes that stable de-Sitter vacua cannot exist in any consistent theory of quantum gravity.
Non-geometric flux compactifications provide potential counter-examples to 
this conjecture, however, in this paper we have argued  that 
all examples known to us  (which only involve 
$H$-, \mbox{$F$-,} $Q$-, $R$- and $F_3$-fluxes)  are  not consistent in string theory. 
On the other hand, we were not able to show in general that stable de-Sitter vacua 
cannot be obtained from non-geometric flux-compactifications. In fact, 
the scalar potential  we derived
shows a close resemblance 
to ordinary Calabi-Yau compactifications 
with the familiar $H$- and $F_3$-flux. 
Whether non-geometric stable de-Sitter vacua exist in string theory remains an open question.

%%%%%%%%%%%%%%%%%%%%%%%%%%%%%%%%%%%%%%%%%%%%%%%
%%%%%%%%%%%%%%%%%%%%%%%%%%%%%%%%%%%%%%%%%%%%%%%

\subsubsection*{Outlook}

The results obtained in this paper provide a starting point for further research 
on non-geometric flux-compactifications:
\begin{itemize}

\item Our requirement from section~\ref{sec_tadpole_charges}  on the tadpole charges $N_A$
has been motivated by plausibility and duality arguments and by extending existing 
swampland conjectures. It would  be desirable to derive 
the condition $N_A\leq 0$ from first principles.

\item For the case $h^{1,1}_{\vphantom{-}}=h^{2,1}_-$ we derived an expression 
for the scalar potential where the R-R three-form flux $F_3$ has been replaced by 
the tadpole charges $N_A$. 
Using our requirement from section~\ref{sec_tadpole_charges},
for a given compactification manifold 
there are finitely-many choices for $N_A$ which in turn simplifies
computer-based searches for flux vacua.

\item The complex-structure moduli sector is subject to a $Sp(2h^{2,1}_-+2,\mathbb Z)$ 
duality acting on the third cohomology \cite{Candelas:1990pi}. 
This implies that the NS-NS $H$-, $F$-, $Q$- and $R$-fluxes contained in the flux matrix 
 \eqref{matrix_002} 
are unique up to \raisebox{0pt}[0pt][0pt]{$Sp(2h^{2,1}_-+2,\mathbb Z)$} transformations. 
As a preliminary study we considered \raisebox{0pt}[0pt][0pt]{$h^{1,1}_{\vphantom{-}}=h^{2,1}_-=1$} 
and flux quanta with values $-3,\ldots,+3$ appearing in \raisebox{0pt}[0pt][0pt]{$\hat\Xi$}. In this case there
are $318\,784$ combinations for which $\hat \Xi$ has maximal rank, 
which by  \raisebox{0pt}[0pt][0pt]{$Sp(2h^{2,1}_-+2,\mathbb Z)$} transformations
are related to $70$ inequivalent choices. 
We therefore have a reduction of the size of the flux landscape as
\eq{
  \mbox{$318\,784$ flux choices} \quad \xrightarrow{\hspace{10pt}Sp(4,\mathbb Z)\hspace{10pt}}\quad 
  \mbox{$70$ inequivalent flux choices} \,.
}
Note however that $Sp(2h^{2,1}_-+2,\mathbb Z)$ also acts on the periods of 
the holomorphic three-form in a non-trivial way and typically does not preserve 
for instance a large-volume limit. 
It would be interesting to extend this observation to more general NS-NS flux configurations.

\end{itemize}
We are planning to come back to these questions in the future.

%%%%%%%%%%%%%%%%%%%%%%%%%%%%%%%%%%%%%%%%%%%%%%%
%%%%%%%%%%%%%%%%%%%%%%%%%%%%%%%%%%%%%%%%%%%%%%%
%%%%%%%%%%%%%%%%%%%%%%%%%%%%%%%%%%%%%%%%%%%%%%%
%%%%%%%%%%%%%%%%%%%%%%%%%%%%%%%%%%%%%%%%%%%%%%%

\vskip1cm
\subsubsection*{Acknowledgments}

We thank
Johan Bl\aa b\"ack,
Cesar Damian,
Adolfo Guarino,
Daniel Junghans,
Oscar Loaiza-Brito and
Thomas van Riet
for helpful discussions and communications. 
The work of EP is supported by a Heisenberg grant of the
\textit{Deutsche Forschungsgemeinschaft} (DFG, German Research Foundation) 
with project-number 430285316.

%%%%%%%%%%%%%%%%%%%%%%%%%%%%%%%%%%%%%%%%%%%%%%%
%%%%%%%%%%%%%%%%%%%%%%%%%%%%%%%%%%%%%%%%%%%%%%%
%%%%%%%%%%%%%%%%%%%%%%%%%%%%%%%%%%%%%%%%%%%%%%%
%%%%%%%%%%%%%%%%%%%%%%%%%%%%%%%%%%%%%%%%%%%%%%%
%%%%%%%%%%%%%%%%%%%%%%%%%%%%%%%%%%%%%%%%%%%%%%%
%%%%%%%%%%%%%%%%%%%%%%%%%%%%%%%%%%%%%%%%%%%%%%%
%%%%%%%%%%%%%%%%%%%%%%%%%%%%%%%%%%%%%%%%%%%%%%%
%%%%%%%%%%%%%%%%%%%%%%%%%%%%%%%%%%%%%%%%%%%%%%%
%%%%%%%%%%%%%%%%%%%%%%%%%%%%%%%%%%%%%%%%%%%%%%%
%%%%%%%%%%%%%%%%%%%%%%%%%%%%%%%%%%%%%%%%%%%%%%%
%%%%%%%%%%%%%%%%%%%%%%%%%%%%%%%%%%%%%%%%%%%%%%%
%%%%%%%%%%%%%%%%%%%%%%%%%%%%%%%%%%%%%%%%%%%%%%%
%%%%%%%%%%%%%%%%%%%%%%%%%%%%%%%%%%%%%%%%%%%%%%%
%%%%%%%%%%%%%%%%%%%%%%%%%%%%%%%%%%%%%%%%%%%%%%%
%%%%%%%%%%%%%%%%%%%%%%%%%%%%%%%%%%%%%%%%%%%%%%%
%%%%%%%%%%%%%%%%%%%%%%%%%%%%%%%%%%%%%%%%%%%%%%%
%%%%%%%%%%%%%%%%%%%%%%%%%%%%%%%%%%%%%%%%%%%%%%%
%%%%%%%%%%%%%%%%%%%%%%%%%%%%%%%%%%%%%%%%%%%%%%%

\clearpage
\nocite{*}
\bibliography{references}
\bibliographystyle{utphys}

%%%%%%%%%%%%%%%%%%%%%%%%%%%%%%%%%%%%%%%%%%%%%%%
%%%%%%%%%%%%%%%%%%%%%%%%%%%%%%%%%%%%%%%%%%%%%%%
%%%%%%%%%%%%%%%%%%%%%%%%%%%%%%%%%%%%%%%%%%%%%%%
%%%%%%%%%%%%%%%%%%%%%%%%%%%%%%%%%%%%%%%%%%%%%%%
%%%%%%%%%%%%%%%%%%%%%%%%%%%%%%%%%%%%%%%%%%%%%%%

\end{document}